\documentclass{article}
\usepackage{amssymb}

\def\un{{\rm 1\mkern-4mu I}}

\def\build#1_#2^#3{\mathrel{
\mathop{\kern 0pt#1}\limits_{#2}^{#3}}}

\def\overcorners{\ulcorner\joinrel\urcorner}
\def\longovercorners{{\ulcorner}\joinrel{\overline{{ \ \atop \ }}}\joinrel{\urcorner}}
\def\llongovercorners{{\ulcorner}\joinrel{\overline{{ \ \ \ \ \atop \ \ \
}}}\joinrel{\urcorner}}

\DeclareSymbolFont{ltrs}     {OML}{cmm}{bx}{it}
\DeclareMathSymbol{\bfzeta}{\mathord}{ltrs}{"10}

\title{Globally conformal invariant gauge field theory with rational
correlation functions\footnote{A brief preview of this paper is contained in
\cite{NST2}.}}
\author{N.M. Nikolov$^{1)}$, Ya. S. Stanev$^{1)2)}$, I.T. Todorov$^{1)3)}$}
\date{
\begin{itemize}
\item[$^{1)}$] {\small Institute for Nuclear Research and Nuclear Energy} \\
{\small Tsarigradsko Chaussee 72, BG-1784 Sofia, Bulgaria}
\item[$^{2)}$] {\small I.N.F.N. -- Sezione di Roma II} \\
{\small Via della Ricerca Scientifica 1, I-00133 Roma, Italy}
\item[$^{3)}$] {\small Section de Math\'{e}matiques, Universit\'{e} de Gen\`{e}ve
\\ 2-4 rue du Li\`{e}vre, cp 240, CH-1211 Gen\`{e}ve, Suisse}
\end{itemize}
}

\begin{document}

\maketitle

\vspace{-260pt}

${}$ \hfill IHES/P/03/16

${}$ \hfill CERN--TH/2003-056

${}$ \hfill ROM2F/2003/12

${}$ \hfill ESI 1335 (2003)

\vspace{220pt}

\begin{abstract}
Operator product expansions (OPE) for the product of a scalar field
with its conjugate are presented as infinite sums of bilocal fields
$V_{\kappa} (x_1 , x_2)$ of dimension $(\kappa , \kappa)$. For a {\it globally
conformal invariant} (GCI) theory we write down the OPE
of $V_{\kappa}$ into a series of {\it twist} (dimension minus rank) $2\kappa$
symmetric traceless tensor fields with coefficients computed from the
(rational) 4-point function of the scalar field.

We argue that the theory of a GCI hermitian scalar field ${\cal L} (x)$ of
dimension 4 in $D = 4$ Minkowski space such that the 3-point functions of a
pair of ${\cal L}$'s and a scalar field of dimension 2 or 4 vanish can be
interpreted as the theory of local observables of a conformally invariant
fixed point in a gauge theory with Lagrangian density ${\cal L}
(x)$.
\end{abstract}

\vspace{4pt}


\vspace{0.1in}


\vspace{0.1in}

\vspace{28pt}

\vglue 20mm

\noindent e-mail addresses: mitov@inrne.bas.bg, stanev@roma2.infn.it,
todorov@inrne.bas.bg

\newpage

\setcounter{section}{-1}

\section{Introduction}\label{sec0}
\setcounter{equation}{0}

The present paper offers a new step in the realization of the
program set up in \cite{NT} and \cite{NST1} of constructing
a $4$--dimensional conformal field theory (CFT) model with
rational correlation functions of observable fields.

Global conformal invariance (GCI) allows to write down
close form expressions for correlation functions
involving (each) a finite number of free parameters.
Thus, the truncated $4$--point function $w_4^t$ of a neutral scalar
field of (integer) dimension $d$ depends on
$\left[ \frac{d^2}{3} \right]$
real parameters ($\left[ a \right]$
standing for the integer part of the positive number $a$).
In addition to GCI it satisfies the constraints of locality
and energy positivity. Operator product expansion (OPE)
provides a method of taking the remaining condition of Wightman
(i.~e. Hilbert space) positivity into account. We organize
systematically OPE into mutually orthogonal bilocal fields
$V_{\kappa} \left( x_1,\, x_2 \right)$ of dimension
$\left( \kappa,\, \kappa \right)$ defined by the condition
that $V_{\kappa}$ can be expanded into an (infinite) sum
(of integrals) of twist $2\kappa$ symmetric traceless
tensor local fields. An algorithm is given for computing the
\textit{full} contribution of each $V_{\kappa}$ to the
$4$--point function. The effectiveness of this approach
is enhanced by the fact that $V_1 \left( x_1,\, x_2 \right)\,$,
which involves an infinite sum of \textit{conserved} tensor
fields, satisfies the d'Alembert equation in each argument and
the $4$--point function
\(\langle 0 \mid V_1 \left( x_1,\, x_2 \right)
V_1 \left( x_3,\, x_4 \right) \mid 0 \rangle\)
is rational by itself. For the model of a $d=2$ neutral scalar
field $\phi \left( x \right)$ the OPE of two $\phi$'s can be
summed up simply as \cite{NST1}
\begin{equation}\label{eq0.1}
\phi \left( x_1 \right) \phi \left( x_2 \right)
\, = \,
\langle 0 \mid \phi \left( x_1 \right)
\phi \left( x_2 \right) \mid 0 \rangle
+ \left( 12 \right) V_1 \left( x_1,\, x_2 \right)
+ : \! \phi \left( x_1 \right) \phi \left( x_2 \right) \! :
\qquad
\end{equation}
where $\left( 12 \right)$ is the free $0$--mass $2$--point
function,
\begin{equation}\label{eq0.2}
\left( 12 \right) \, = \, \frac{1}{4 \pi^2 \rho_{12}}
\, , \quad
\rho_{12} \, = \, x_{12}^{\, 2} + i \, 0 \, x_{12}^0
\, , \quad
x_{12} \, = \, x_1 - x_2
\, , \quad
x^2 \, = \, \mathbf{x}^2 - x_0^2
\, , \qquad
\end{equation}
and the \textit{normal product}
$: \! \phi \left( x_1 \right) \phi \left( x_2 \right) \! :$
defined by (\ref{eq0.1}) is non--singular for \(x_1 = x_2\).
The simplicity of $V_1$ in this model has been exploited
in \cite{NST1} to prove that it can be written as a sum of
normal products of (mutually commuting) free $0$--mass fields.

We focus in the present paper on the study of a model
generated by a (neutral) scalar field which can be interpreted
as a (gauge invariant)
Lagrangian density ${\cal L} (x)$. Taking into account the above cited
triviality result for $d=2$ we require that the OPE of ${\cal L} (x_1) {\cal
L} (x_2) - \langle 0 \mid {\cal L} (x_1) {\cal L} (x_2) \mid 0 \rangle$ does
not involve any field of dimension lower than 4 (which just amounts to
excluding a possible scalar field of dimension 2). This requirement
decreases the number of free parameters in $w_4^t$ from five to four.
It already excludes most of
the standard renormalizable interaction Lagrangians (like Yukawa and
${\varphi}^4$) but allows for the Lagrangian ${\cal L}$ and for the
pseudoscalar topological term $\widetilde{{\cal L}}$ of a pure gauge theory
\begin{eqnarray}
\label{eq1.11}
{\cal L} (x) \, = && \hspace{-15pt}
- \frac{1}{4} \, tr (F_{\mu \nu} (x) \, F^{\mu \nu} (x)) \, ,
\\ \label{eq1.11add}
\widetilde{{\cal L}} (x) \, = && \hspace{-15pt}
\frac{1}{4} \, \varepsilon^{\kappa\lambda\mu\nu} \,tr (
F_{\kappa\lambda} (x) \, F_{\mu\nu} (x)) \, .
\end{eqnarray}
The invariance of ${\cal L}$ under
conformal rescaling of the metric is made manifest by
writing the action density in terms of the Yang-Mills curvature form
\begin{equation}
\label{eq1.12}
F(x) := \frac{1}{2} \, F_{\mu \nu} (x) \, dx^{\mu} \wedge dx^{\nu}
\end{equation}
and its Hodge dual, $^*F$:
\begin{equation}
\label{eq1.13}
{\cal L} (x) \, \sqrt{\vert g \vert} \, dx^0 \wedge dx^1 \wedge dx^2 \wedge
dx^3 = tr (^*F (x) \wedge F(x))
\end{equation}
(where $\vert g \vert$ is the absolute value of the determinant of the
metric tensor, $\vert g \vert = 1$ for the Minkowski metric). (It does not
seem superfluous to reiterate that the conformal invariance of the gauge
field Lagrangian singles out $D=4$ as the dimension of space time \cite{C}.)

It seems appropriate to demand further invariance of the theory under
``electric-magnetic duality''\footnote{The authors thank Dirk Kreimer for
this suggestion.}, -- i.e., under the change $F \rightarrow \, ^*F$. Taking
into account the fact that the Hodge star defines a complex structure in
Minkowski space, we deduce that the Lagrangian (\ref{eq1.13}) changes sign
under $*$:
\begin{equation}
\label{eq1.14}
^*(^*F) = -F \Rightarrow {\cal L} (^*F) = - {\cal L} (F) \, .
\end{equation}
It follows that odd point correlation functions of ${\cal L} (x)$ should
vanish.
Requiring space reflection
invariance we also find that the vacuum expectation values of an odd
number of $\widetilde{{\cal L}}$'s (and any number of ${\cal L}$'s)
should vanish. This leads us to the assumption that no (pseudo) scalar
field of dimension 4 appears in the OPE of two ${\cal L}$'s thus eliminating
one more parameter in the expression for
$w_4^t$.

The case of a non-abelian gauge field is distinguished by the existence
of a non-trivial 3-point function of $F$ which agrees with local
commutativity of Bose fields. Consistency of the equations of motion
satisfied by the 2- and 3-point functions of F with OPE, however,
requires the use of indecomposable representations of the conformal group
${\cal C}$ which makes unpractical exploiting the compositeness of
${\cal L}$ (Sec.~4.2).

The paper is organized as follows.

We outline  basic ideas and review earlier work and notation
in Sec.~1. In particular, we make precise the notation of an
elementary positive energy representation of ${\cal C}$,
and indicate an extension of the present approach to any even
space-time dimension $D$.

Sec.~\ref{ssec2.1} provides a general treatment of the OPE of a pair of
conjugate scalar fields of dimension $d \in {\mathbb N}$ in terms of
bilocal fields $V_{\kappa} (x_1 , x_2)$. It also reproduces the formula for
the expansion of $V_{\kappa}$ into an infinite series of symmetric traceless
tensor fields of twist $2\kappa$.
Sec.~\ref{ssec2.2} displays the (crossing symmetrized) contribution of twist
2 ({\it conserved}) tensors and their light cone expansions for $d=2$ and
$d=4$.

In Sec.~\ref{ssec3.1} we study systematically the truncated 4-point function
of the $d=4$ field ${\cal L} (x)$ and discuss the special case of the
Lagrangian ${\cal L}_0$ (\ref{eq3.9}) of the free Maxwell field.
In Sec.~3.~2 we analyse the conformally invariant
(rational) factor $f_1$ of the harmonic 4-point function
\begin{equation}\label{eq1.18new}
\langle 0 \mid V_{1}(x_1,x_2) V_{1}(x_3,x_4) \mid 0 \rangle \, = \,
(13)(24)f_{1}(s,t) \, , \
\end{equation}
displaying a basis of solutions $j_{\nu}(s,t)$ of the ``conformal Laplace
equation'' (\ref{eq3.7new}) which admits a natural crossing symmetrization.
Sec~\ref{ssec3.2} displays the operator content and the light cone expansion
in twist 4 and twist 6 tensor fields in the case of general~${\cal L}$.

The restrictions on the parameters of the truncated 4-point function
coming from Wightman positivity of the singular part of the the s-channel
OPE are analysed in Sec.~4.1.
Sec.~\ref{sec4} is devoted to a summary of results and concluding remarks.

Appendix~\ref{apA} summarizes the derivation of the main result of \cite{NST1}
referred to in the text discussing on the way the possibility for a similar
treatment of the $d=4$ case of interest (and the difficulties lying ahead).

\section{Consequences of GCI (a synopsis)}\label{sec1}
\setcounter{equation}{0}

\subsection{Elementary positive energy representations of the conformal
group}\label{ssec1.1}

The idea that quantum fields should be subdivided into local observables and
``charged fields'' (that are relatively local to the observables and act as
intertwiners among different superselection sectors) emerged from several
decades of work of Haag and collaborators reviewed in \cite{H}. It has been
implemented in 2-dimensional conformal field theory (CFT) models in which
chiral currents (including the stress-energy tensor) play the part of local
observables while primary (with respect to a given chiral algebra) ``chiral
vertex operators'' correspond to ``charged fields'' intertwining between
different superselection sectors. (Among the numerous reviews on 2D CFT we
cite the textbook \cite{DMS} and the earlier article \cite{FST} which refers
to the axiomatic quantum field theory framework.) Local
observable fields are assumed to satisfy Wightman axioms \cite{SW}. More
specifically, we demand that their set contains both the stress energy tensor
$T_{\mu \nu} (x)$ (cf.~\cite{MS}) and the Lagrangian density ${\cal
L} (x)$.

To set the stage we fix the class of local field representations of the
quantum mechanical conformal group ${\cal C} (D) = {\rm Spin} (D,2)$ under
consideration (cf.~\cite{DMPPT}, \cite{M}, \cite{T1}).
The forward tube,
\begin{equation}\label{eq1.1new}
\mathfrak{T}_+ = \left\{ x+iy \, ; \, y^0 > |\mathbf{y}| :=
\left( y_1^2 + \dots + y_{D-1}^2 \right)^{\frac{1}{2}} \right\}
\, , \qquad
\end{equation}
the primitive analyticity domain of the vector valued function
$\psi \left( x+iy \right) \mid 0 \rangle$ for any Wightman field $\psi$,
appears as a homogeneous space of ${\cal C} \left( D \right)$,
\begin{equation}\label{eq1.2new}
\mathfrak{T}_+ = {\cal C} \left( D \right) / K \left( D \right)
\quad \mathrm{for} \quad
K \left( D \right) = \mathrm{Spin} \left( D \right) \times
U \left( 1 \right)
\, , \qquad
\end{equation}
$K \left( D \right)$ being the maximal compact subgroup of
${\cal C} \left( D \right)\,$.
An \textit{elementary positive energy representation} of
${\cal C} \left( D \right)$ is one induced by a (finite
dimensional) irreducible representation of $K \left( D \right)$.
Any such representation gives rise to a transformation law
for local fields defined (as operator valued distributions)
on the (conformally) compactified Minkowski space $\overline{M}_D$
which appears as a (part of the) boundary of $\mathfrak{T}_+\,$.
$\overline{M}_D$ also appears as a homogeneous
space of ${\cal C} (D)$ with respect to a $\left( {D+1 \atop 2} \right) + 1$
parameter ``parabolic subgroup'' ${\cal H} (D) \subset {\cal C} (D)$ defined
as a semidirect product ${\cal H} (D) = {\cal N}_D \rtimes ({\rm Spin} (D-1
, 1) \times SO(1,1))$ of a $D$ dimensional abelian subgroup ${\cal N}_D$ of
``special conformal transformations'' with the direct product of the
(quantum mechanical) Lorentz group and the 1-parameter subgroup of ${\cal C}
(D)$ of uniform dilations:
\begin{equation}
\label{eq1.1}
\overline{M}_D \simeq {\cal C} (D) / {\cal H} (D) \simeq ({\mathbb S}^1
\times {\mathbb S}^{D-1}) / {\mathbb Z}_2 \, .
\end{equation}
The resulting \textit{elementary} (\textit{local field})
\textit{representation} \cite{T1} can be defined as a
positive energy representation of ${\cal C} (D)$ induced by a finite
dimensional irreducible representation of ${\cal H} (D)$ (energy positivity
fixing the representation of the discrete centre -- ${\mathbb Z}_4$ for even
$D$ -- of ${\cal C} (D)$ -- see \cite{M}). In the case $D=4$ of interest
(studied in \cite{M}),
\begin{equation}
\label{eq1.2}
{\cal C} \equiv {\cal C} (4) = SU(2,2) (\cong {\rm Spin} (4,2)) \, , \
\overline M (\equiv \overline M_4) = U(2) (= ({\mathbb S}^1 \times {\mathbb
S}^3) / {\mathbb Z}_2)
\, ,
\end{equation}
the elementary representations of ${\cal C}$ are labeled by triples
$(d,j_1,j_2)$ of non-negative half integers,
which admit two equivalent to each other interpretations.
Viewed as labels of a $K$--induced representation $d$ stands for the
minimal eigenvalue of the conformal Hamiltonian (defined below)
while $\left( j_1,\, j_2 \right)$ label the irreducible representations
of $SU \left( 2 \right) \times SU \left( 2 \right)$.
Alternatively, referring to a ${\cal H}$--induced representation,
$d$ is the conformal
dimension (which may take all positive values if we substitute ${\cal C}$ by
its universal covering) and $(j_1 , j_2)$ stands for a $(2j_1 + 1)(2j_2 +
1)$ dimensional representation of the Lorentz group (a spin-tensor of $2j_1$
undotted and $2j_2$ dotted indices).
The elementary positive energy representation of ${\cal C}$ are,
in general, neither unitary nor irreducible.
They are proven \cite{M} to be unitary (or to admit an
irreducible unitary subrepresentation) iff
\begin{eqnarray}
\label{eq1.2bis}
\hbox{(a)} &d \geq 1 + j_1 + j_2 \qquad &\hbox{for} \ j_1 j_2 = 0 \, ,
\hfill \nonumber \\
\hbox{(b)} &d \geq 2 + j_1 + j_2 \qquad &\hbox{for} \ j_1 \, j_2 > 0 \, .
\end{eqnarray}
For a symmetric traceless tensor with $\ell$
indices (i.e. for $2j_1=2j_2=\ell$) the counterpart of (\ref{eq1.2bis})
can be readily written for any space-time dimension $D$:
\begin{equation}
\label{1.3D}
d \, \geq \, d_0 \, := \, \frac{D-2}{2}
\quad \hbox{for} \ \ell \, = \, 0;\qquad d \, \geq \, D-2+\ell
\quad \hbox{for} \ \ell \, = \,1,\, 2,\, \dots\, ,
\end{equation}
$d_0$ being the dimension of a free 0--mass scalar field.
It is quite remarkable that conventional relativistic (quantum) fields fit
precisely the above definition (rather than transforming under Wigner's
unitary irreducible representations of the Poincar\'e group --
cf.~\cite{BLOT} or \cite{SW}). Furthermore, a free scalar, Weyl spinor and
Maxwell tensor correspond to the lower limit of the above series (a),
for weights $(1,0,0)$, $\left( \frac{3}{2} , \frac{1}{2} , 0
\right)$ and $(2,1,0) + (2,0,1)$, respectively, while conserved symmetric
traceless tensors (starting with the vector current $\left( 3 , \frac{1}{2}
, \frac{1}{2} \right)$) are {\it twist two fields} (fitting the lower limit
of the series (b) in (\ref{eq1.2bis}) and of (\ref{1.3D})).

In all the latter cases it is the subspace of solutions of the
corresponding field equations or conservation law that
carries a unitary representation of ${\cal C}$.

Note further that the {\it conformal Hamiltonian}, the generator of the
centre, $U(1)$, of the maximal compact subgroup,
${\rm Spin} (D) \times U(1) \, / \, \mathbb{Z}_2$
of ${\cal C} (D)$ ($S(U(2) \times U(2))$ for $D=4$) is positive
definite whenever the Minkowski space energy is (cf.~\cite{S}) and has a
discrete spectrum belonging to the set $\{ d+n \, ; \ n=0,1,2,\ldots \}$ for
the positive energy representation of ${\cal C}$ of weight $(d,j_1 , j_2)$.

\subsection{Huygens principle and rationality of conformally invariant
correlation functions}\label{ssec1.2}

Global conformal invariance (GCI) and local commutativity in four
dimensional (4D) Minkowski space $M$ imply the Huygens principle which
can be stated in the following strong form \cite{NT}. Let $\psi (x)$ be a
GCI local (Bose or Fermi) field of weight $(d,j_1,j_2)$; then $d+ j_1 + j_2$
should be an integer and
\begin{equation}
\label{eq1.3}
(x_{12}^2)^{d+j_1+j_2} (\psi (x_1) \, \psi^* (x_2) -(-1)^{2j_1 + 2j_2} \,
\psi^* (x_2) \, \psi (x_1)) = 0
\end{equation}
for {\it all} $x_1 , x_2$ in $M$. The Huygens principle and energy
positivity (more precisely, the relativistic spectral conditions) imply
rationality of correlation functions (\cite{NT}, Theorem~3.1).

Wightman positivity in 4D restricts the degree of singularities of {\it
truncated} ($\equiv$ connected) $n$-point functions (which can only occur
for coinciding arguments): they must be strictly lower than the degree of
the pole of the corresponding 2-point propagator. This allows to determine
any correlation function up to a finite number of (constant) parameters. In
particular, the truncated 4-point function of a hermitian scalar field
$\phi$ of (integer) dimension $d \in {\mathbb N}$ can be written in the form
(\cite{NST1} Sec.~1)
\begin{eqnarray}
\label{eq1.4}
w_4^t &\equiv &w^t (x_1 , x_2 , x_3 , x_4) := \langle 1234 \rangle - \langle
12 \rangle \langle 34 \rangle - \langle 13 \rangle \langle 24 \rangle -
\langle 14 \rangle \langle 23 \rangle \nonumber \\
&= &
(12)(23)(34)(14)
\left( \frac{\rho_{13} \rho_{24}}{
\rho_{12} \rho_{23} \rho_{34} \rho_{14}} \right)^{d-2}
\, P_d (s,t)
\end{eqnarray}
where
\begin{equation}
\label{eq1.5}
\langle 1 \ldots n \rangle := \langle 0 \mid \phi (x_1) \ldots \phi (x_n)
\mid 0 \rangle \, ,
\end{equation}
$\left( 12 \right)$ is the free massless propagator (\ref{eq0.2}),
$s$ and $t$ are the conformally invariant cross-ratios (so that the
prefactor in (\ref{eq1.4}) is a rational function of $\rho_{ij}$):
\begin{equation}
\label{eq1.6}
s = \frac{\rho_{12} \rho_{34}}{\rho_{13} \rho_{24}} \, , \ t =
\frac{\rho_{14} \rho_{23}}{\rho_{13} \rho_{24}} \, ,
\end{equation}
$\rho_{ij}$ are defined in (\ref{eq0.2})
and $P_d (s,t)$ is a polynomial in $s$ and $t$ of total degree $2d-3$
(\(P_1 \left( s,\, t \right) \equiv 0\)):
\begin{equation}
\label{eq1.7}
P_d (s,t) = \sum_{{i,j \geq 0 \atop i+j \leq 2d-3}} c_{ij} \, s^i \, t^j \,
.
\end{equation}
Furthermore, local commutativity implies the crossing symmetry conditions
\begin{equation}
\label{eq1.8}
s_{12} \, P_d (s,t) = P_d (s,t) = s_{23} \, P_d (s,t)
\end{equation}
where $s_{ij}$ is the substitution exchanging the arguments $x_i$ and $x_j$:
$$
s_{12} \, P_d (s,t) := t^{2d-3} \, P_d \left( \frac{s}{t} , \frac{1}{t}
\right) \, ,
$$
\begin{equation}
\label{eq1.9}
s_{23} \, P_d (s,t) := s^{2d-3} \, P_d \left( \frac{1}{s} , \frac{t}{s}
\right) \, .
\end{equation}
Invariance under (\ref{eq1.9}) implies the symmetry property $s_{13} \, P_d
(s,t) := P_d (t,s) = P_d (s,t)$ where $s_{13} = s_{12} s_{23} s_{12} =
s_{23} s_{12} s_{23}$. (The Wightman function $w_4$ of a neutral scalar
field is in fact symmetric -- as a rational function -- under the group
${\cal S}_4$ of permutations of the four position variables $x_a$, $a =
1,2,3,4$. The normal subgroup ${\mathbb Z}_2 \times {\mathbb Z}_2$ of ${\cal
S}_4$, however, generated by $s_{12} s_{34}$ and $s_{14} s_{23}$ leaves the
conformal cross ratios $s$ and $t$ invariant, so that only the factor group
${\cal S}_4 / {\mathbb Z}_2 \times {\mathbb Z}_2 \simeq {\cal S}_3$ acts
effectively on $P_d (s,t)$.) The number of independent crossing symmetric
polynomials is $\left[ \frac{d^2}{3} \right]$ (the integer part of
$\frac{d^2}{3}$: $1$ for $d=2$, $3$ for $d=3$, $5$ for $d=4$).

\section{OPE in terms of bilocal fields}\label{sec2}
\setcounter{equation}{0}

\subsection{Bilocal fields as infinite series of symmetric tensor fields of a
given twist}\label{ssec2.1}

It appears convenient, at least in analysing a (truncated) 4-point function,
to organize the infinite series of integrals of local tensor fields in an OPE
into a finite sum of bilocal fields. In order to avoid purely, technical
complications we shall exhibit the basic idea in the simplest case of the
product of a $d$-dimensional {\it scalar field} $\psi (x)$ (in $M=M_4$) with
its conjugate. We look for an expansion of the form
\begin{equation}
\label{eq2.1}
\psi^* (x_1) \, \psi (x_2) = \langle 12 \rangle + \sum_{\kappa = 1}^{d-1}
(12)^{d-\kappa} \, V_{\kappa} (x_1 , x_2) + : \psi^* (x_1) \, \psi (x_2) : \,
,
\end{equation}
where $(12)$ is the free massless propagator defined in (\ref{eq0.2}),
$\langle 12 \rangle$ is the 2-point function
\(
\langle 12 \rangle = \langle 0 \mid \psi^* (x_1) \, \psi (x_2) \mid 0 \rangle
= N_{\psi} (12)^d \, ,
\)
and $V_{\kappa} (x_1 , x_2)$ is a bilocal conformal field of dimension
$(\kappa , \kappa)$ which is assumed to have an expansion in a series of
local (hermitian) symmetric traceless tensor fields
\begin{equation}
\label{eq2.3}
O_{2\kappa , \ell} (x ; \zeta) = O_{2\kappa , \mu_1 \ldots \mu_{\ell}} (x)
\, \zeta^{\mu_1} \ldots \zeta^{\mu_{\ell}} \, , \ \Box_{\zeta} \, O_{2\kappa
\ell} (x;\zeta)  = 0
\end{equation}
(for $\Box_{\zeta} := \frac{\partial^2}{\partial {\bfzeta}^2} -
\frac{\partial^2}{\partial \zeta_0^2}$) of dimension $2\kappa + \ell$ (i.e.
of {\it fixed twist} $2\kappa$). We shall make use of the fact that the
harmonic polynomial $O_{2\kappa , \ell} (x,\zeta)$ in the auxiliary
variable $\zeta$ is uniquely determined by its values on the light cone
$\zeta^2 = 0$ \cite{BT}. Whenever such an expansion is valid, it can be
written quite explicitly\footnote{Covariant conformal OPE were proposed in
\cite{FGGP}, soon after the pioneer work of Wilson \cite{W}. Further
developments are reviewed in \cite{DMPPT}, \cite{TMP}, \cite{OP}, and
\cite{FP}; we found useful the recent presentation \cite{DO}.}:
\begin{equation}
\label{eq2.4}
V_{\kappa} (x_1 , x_2) = \sum_{\ell = 0}^{\infty} C_{\kappa \ell} \int_0^1
K_{\kappa \ell} (\alpha , \rho_{12} \, \Box_2) \, O_{2\kappa , \ell}
(x_2 + \alpha x_{12} ; x_{12}) \, d\alpha
\end{equation}
where
$$
K_{\kappa \ell} (\alpha , z) = \sum_{n = 0}^{\infty} \frac{[\alpha
(1-\alpha)]^{\ell + \kappa + n-1}}{B(\ell + \kappa , \ell + \kappa)}
\frac{\left( - \frac{z}{4} \right)^n}{n!(\ell + 2\kappa -d_0)_n}
$$
\begin{equation}
\label{eq2.5}
\left( (\lambda)_n = \frac{\Gamma (\lambda + n)}{\Gamma (\lambda)} \right) \, ,
\end{equation}
$B$ is the Euler beta function, $\Gamma (\mu + \nu) \, B (\mu , \nu) = \Gamma
(\mu) \, \Gamma (\nu)$, and the d'Alembert operator $\Box_2$ acts on $x_2$
for fixed $x_{12}$.
The normal product in (\ref{eq2.1}) has a similar expansion
\begin{equation}
\label{eq2.5new}
: \! \psi^* \left( x_1 \right) \psi \left( x_2 \right) \! :
\, = \,
\sum_{\kappa \, = \, d}^{\infty}
\rho_{12}^{\kappa -d} \, V_{\kappa} \left( x_1,\, x_2 \right)
\, \qquad
\end{equation}
thus involving all higher (even) twists.

\medskip

{\it Remark 2.1.}
A field $V(x_1,x_2)$ is called {\it bilocal} if it commutes with
any local field ${\phi}(x_3)$ for spacelike $x_{13}$ and $x_{23}$.
We note that GCI does not imply the Huygens principle for bilocal fields;
hence, their correlation functions need not be rational.
It is noteworthy that
for any even space-time dimension $D$,
$V_{d_0}$ does have rational correlation functions
as will be demonstrated in Sec. 2.2 below using conservation of twist
$2{d_0}$ tensors.
We shall also see (in Sec.~3.~3) that this is not the case for
$V_{\kappa}$ if $\kappa>d_0$.

In order to verify (\ref{eq2.4}), (\ref{eq2.5}) we need the general
conformally invariant expression for the 2- and 3-point functions \cite{TMP}
$$
\langle 0 \mid O_{2\kappa , \ell} (x_1 , \zeta_1) \, O_{2\kappa , \ell'} (x_2
, \zeta_2) \mid 0 \rangle = \delta_{\ell\ell'} N_{\kappa \ell} (12)^{2\kappa}
(\zeta_1 R (x_{12}) \zeta_2)^{\ell}
$$
\begin{equation}
\label{eq2.6}
\hbox{for} \ \zeta_{1,2}^2 = 0 \, ,
\end{equation}

\begin{equation}
\label{eq2.7}
\langle 0 \mid V_{\kappa} (x_1 , x_2) \, O_{2\kappa , \ell} (x_3 , \zeta)
\mid 0 \rangle = A_{\kappa \ell} (13)^{\kappa} (23)^{\kappa} (X_{12}^3
\zeta)^{\ell}
\end{equation}
where we have introduced the symmetric tensor
\begin{equation}
\label{eq2.8}
\rho_{12} \zeta_1 R (x_{12}) \zeta_2 = \zeta_1 r (x_{12}) \zeta_2 := \zeta_1
\zeta_2 - 2 \frac{(\zeta_1 x_{12}) (\zeta_2 x_{12})}{\rho_{12}} \quad
(r(x)^2 = \un)
\end{equation}
and the vector
\begin{equation}
\label{eq2.9}
X_{12}^3 = \frac{x_{13}}{\rho_{13}} - \frac{x_{23}}{\rho_{23}} \ \hbox{of
square} \ (X_{12}^3)^2 = \frac{\rho_{12}}{\rho_{13} \rho_{23}} \, .
\end{equation}
The integrodifferential operator in the right hand side of (\ref{eq2.4}) is
characterized by the property to transform a 2-point function of type
(\ref{eq2.6}) into a 3-point one, (\ref{eq2.7}) (see \cite{DO}):
\begin{equation}
\label{eq2.10}
\int_0^1 d \alpha K_{\kappa\ell} (\alpha , \rho_{12} \Box_2) (x_{12} R
(y(\alpha)) \zeta)^{\ell} \rho_{y(\alpha)}^{-2\kappa} = \frac{(X_{12}^3
\zeta)^{\ell}}{(\rho_{13} \rho_{23})^{\kappa}} \ \hbox{for} \ \zeta^2 = 0
\, ,
\end{equation}
where $y(\alpha) = x_{23} + \alpha x_{12} = \alpha x_{13} + (1-\alpha) x_{23}$,
$\rho_y = y^2 + i 0 y^0$. The above relations are particularly easy to verify
for $\rho_{12} = 0$ as the second argument of $K_{\kappa \ell}$ then
vanishes and we obtain the light-cone expansion of $V_{\kappa}$.

Eqs.~(\ref{eq2.4}), (\ref{eq2.6}) and (\ref{eq2.10}) imply that the constants
$A_{\kappa \ell}$ in (\ref{eq2.7}) are proportional to the expansion
coefficients $C_{\kappa \ell}$ in (\ref{eq2.4}). Furthermore, a simple
analysis shows that only their product is invariant under rescaling of
$O_{2\kappa , \ell}$ (for fixed $V_{\kappa}$).

We emphasize that the expansion (\ref{eq2.4}) is universal: only the
coefficients $C_{\kappa \ell}$ depend on the field $\psi$ in (\ref{eq2.1}).

It follows from (\ref{eq2.1}) that each $V_{\kappa}$ satisfies the symmetry
condition
\begin{equation}
\label{eq2.11}
[V_{\kappa} (x_1 , x_2)]^* = V_{\kappa} (x_2 , x_1) \, ;
\end{equation}
taking into account the reality of $O_{2\kappa , \ell}$ we deduce
\begin{equation}
\label{eq2.12}
C_{\kappa\ell}^* = (-1)^{\ell} C_{\kappa\ell} \, .
\end{equation}
If $\psi$ is hermitian then so is $V_{\kappa}$, and $C_{\kappa 2\ell + 1}$
vanish:
\begin{equation}
\label{eq2.13}
C_{\kappa\ell} = 0 \ \hbox{for odd} \ \ell \ \hbox{if} \ V_{\kappa} (x_1 ,
x_2) = V_{\kappa} (x_2 , x_1) \quad (\psi^* = \psi) \, .
\end{equation}

The tensor fields $O_{2\kappa ,\ell}$ transform under elementary
representations of ${\cal C}$. It is convenient to work with such conformal
fields, since they are mutually orthogonal under vacuum expectation values
(see, e.g., \cite{T1} and references therein). It follows that each term in
the expansion (\ref{eq2.1}) is orthogonal to all others:
$$
\langle 0 \mid V_{\kappa} \mid 0 \rangle = 0 = \langle 0 \mid V_{\kappa} (x_1
, x_2) V_{\lambda}^* (x_3 , x_4) \mid 0 \rangle \ \hbox{for} \ \kappa \ne
\lambda \, ,
$$
\begin{equation}
\label{eq2.14}
\langle 0 \mid V_{\kappa} (x_1 , x_2) : \psi (x_3) \psi^* (x_4) : \mid 0
\rangle = 0 \, ,
\end{equation}
as the normal product (\ref{eq2.5new}) is expanded in higher twist fields.

From now on we restrict attention to the study of the OPE algebra of a {\it
hermitian} GCI scalar field of dimension $d$.

For a given (rational) truncated 4-point function we can compute the
invariant under rescaling products
\begin{equation}
\label{eq2.15}
B_{\kappa\ell} := A_{\kappa 2\ell} \, C_{\kappa 2\ell}
\end{equation}
(\(
\hbox{for} \ O_{2\kappa 2\ell} \rightarrow \lambda O_{2\kappa 2\ell} , \,
A_{\kappa 2\ell} \rightarrow \lambda A_{\kappa 2\ell} , \, C_{\kappa 2\ell}
\rightarrow \lambda^{-1}  C_{\kappa 2\ell} , \, B_{\kappa\ell}
\rightarrow B_{\kappa\ell}
\)
with $A$ and $C$ introduced in (\ref{eq2.7}), (\ref{eq2.4})) using the light
cone expansion of $V_{\kappa} (x_3 , x_4)$ in the 4-point function
\begin{equation}
\label{eq2.16}
\langle 0 \mid V_{\kappa} (x_1 , x_2) V_{\kappa} (x_3 , x_4) \mid 0 \rangle =
(13)^{\kappa} (24)^{\kappa} f_{\kappa} (s,t) \, .
\end{equation}
As a consequence of (\ref{eq2.13}) the conformally invariant amplitude
$f_{\kappa}$ is $s_{12}$ symmetric:
\begin{equation}
\label{eq2.17}
s_{12} f_{\kappa} (s,t) := t^{-\kappa} f_{\kappa} \left( \frac{s}{t} , \frac{1}{t}
\right) = f_{\kappa} (s,t) \, .
\end{equation}
Combining (\ref{eq2.4}), (\ref{eq2.7}), (\ref{eq2.15}) and using the standard
integral representation for the hypergeometric function we find
\begin{equation}
\label{eq2.18}
f_{\kappa} (0,t) = \sum_{\ell = 0}^{\infty} B_{\kappa\ell}
(1-t)^{2\ell} F (2\ell + \kappa, 2\ell + \kappa; 4\ell + 2\kappa; 1-t) \, .
\end{equation}
The symmetry condition (\ref{eq2.17}) is reflected in a known transformation
formula for the hypergeometric function:
\begin{equation}
\label{eq2.19}
F \left( a,a;2a; \frac{z}{z-1} \right) = (1-z)^a F (a,a;2a;z) \ \hbox{for} \
a=2\ell + \kappa \, , \ z=1-t \, .
\end{equation}

\subsection{Crossing symmetrized contribution of conserved \\
tensors}\label{ssec2.2}

The case $\kappa = d_0$ (\ref{1.3D}) is of particular interest since the expansion then
comprises all conserved tensors $O_{2d_0,\ell} (x;\zeta) =: T_{\ell} (x,\zeta)$
($T_2$ being the stress-energy tensor). The 3-point functions (\ref{eq2.7})
are all harmonic in $x_1$ and $x_2$; hence, so is $V_{d_0}$:
\begin{equation}
\label{eq2.20}
\Box_1 V_{d_0} (x_1 , x_2) = 0 = \Box_2 V_{d_0} (x_1 , x_2) \, .
\end{equation}
Applying ${\Box}_1$ to both sides of (\ref{eq2.16}) for $\kappa=d_0$ we find
\begin{equation}
\label{new2.21}
\Box_1 \left\{
\left( 13 \right)^{d_0} \left( 24 \right)^{d_0} \,
f_{d_0} \left( s,\, t \right) \right\} \, = \,
\frac{4st}{\rho_{12}\,\rho_{14}} \ \rho_{24} \,
\left( 13 \right)^{d_0} \left( 24 \right)^{d_0}
\Delta \, f_{d_0} \left( s,\, t \right) \, = \, 0
\, \qquad
\end{equation}
where
\begin{equation}
\label{eq2.21}
\Delta = s \frac{\partial^2}{\partial s^2} + t
\frac{\partial^2}{\partial t^2} + (s+t-1) \frac{\partial^2}{\partial s \partial t} +
\frac{D}{2}
\left( \frac{\partial}{\partial s} + \frac{\partial}{\partial t} \right) \, .
\end{equation}
For $D=4$ the operator ${\Delta}$ has been introduced in \cite{EPSS} (in
the context of $N=4$ supersymmetric Yang-Mills theory\footnote{The
authors thank Emery Sokatchev for this remark.}). The general
solution of (\ref{new2.21}) can be written in terms of the chiral variables
$u$ and $v$
exploited in \cite{DO} (a similar procedure is used in Sec. 7 of \cite{EPSS};
see also Appendix A to \cite{NST1}). It is given by
\begin{equation}
\label{eq2.22new}
f_{d_0} (s,t) = \frac{g (u) - g (v)}{\left( u - v \right)^{d_0}}
\quad \hbox{for}
\quad s = u v \, , \ t = (1-u)(1-v) \, .
\end{equation}
(For euclidean $x_j$ the variables $u$ and $v$ are complex conjugate
to each other.)
For a $d$ dimensional field ($d>d_0$) we are looking for a solution
\(f = f_{d_{0}}\) of ${\Delta}f=0$ (\ref{new2.21}) such that the product
of $t^{d-1}f$ is a polynomial in
$s$ and $t$ of overall degree not exceeding $2d-3$ (as a
consequence of (\ref{eq1.7})). It can be obtained by viewing
Eq.~(\ref{new2.21}) as a Cauchy problem with initial condition (satisfying the
symmetry property (\ref{eq2.17}))
\begin{equation}
\label{eq2.22}
f (0,t) = (1+t^{-d_0}) \sum_{\nu = 0}^{d-d_0+1} a_{\nu} \,
\frac{(1-t)^{2\nu}}{t^{\nu}}
\left( = t^{-d_0} f \left( 0 , \frac{1}{t} \right)\right) \, .
\end{equation}
It thus depends on $d-d_0$ parameters $a_0 , a_1 , \ldots , a_{d-d_0-1}$.
Noting that for $s=0=v$ we have $t=1-u$, we can write the solution in the
form (\ref{eq2.22new}) with
\begin{equation}
\label{new2.25}
u^{-d_0} \, g \left( u \right)\, = \, f \left( 0,\, 1-u \right)
\, = \, \left[ 1 \, + \, \left( 1-u \right)^{-d_0} \right]
\sum_{\nu = 0}^{d-d_0+1} a_{\nu} \,
\frac{u^{2\nu}}{\left( 1-u \right)^{\nu}}
\, . \qquad
\end{equation}

\medskip

\textbf{Proposition 2.1.}
\textit{The equation}
\begin{equation}
\label{new2.26}
\Box_1 \, \int_0^1 \!\!
K_{d_0,\hspace{1pt} \ell}
(\alpha ,\hspace{1pt} \rho_{12} \, \Box_y)
\
O_{2d_0 ,\hspace{1pt} \ell} \hspace{-1pt}
(y ;\hspace{1pt} x_{12})
\ d\alpha \, = \, 0
\quad \hbox{\textit{for}} \quad y \, = \, x_2 \, + \, \alpha x_{12}
\, \qquad
\end{equation}
\textit{where} $K_{{\kappa}\ell}$ \textit{is given by} (\ref{eq2.5})
\textit{is necessary and sufficient for the
conservation of the (traceless) twist} $2d_0$ \textit{tensor}
$O_{2d_{0}\ell}$:
\begin{equation}
\label{new2.27}
\frac{\partial^2}{\partial y \, \partial x_{12}} \
O_{2d_0 ,\hspace{1pt} \ell} \hspace{-1pt}
(y ;\hspace{1pt} x_{12}) \, = \, 0
\, . \qquad
\end{equation}

\medskip

\textit{Sketch of proof.} We shall verify that (\ref{eq2.5}) and (\ref{new2.27})
imply (\ref{new2.26}). (The
proof of the sufficiency of (\ref{new2.26}) for the validity of (\ref{new2.27})
uses the same computation.)
The statement follows from the differentiation formula
\begin{eqnarray}
\label{new2.28}
\Box_1 \left\{ \rho_{12}^n \, O_{2d_0 ,\hspace{1pt} \ell} \hspace{-1pt}
(y ;\hspace{1pt} x_{12}) \right\} \, = \,
\rho_{12} \left\{\raisebox{18pt}{}\right.
&& \hspace{-19pt}
4n \left( n+\ell + d_0 +
\alpha\ \frac{\partial}{\partial {\alpha}} \right)
\, + \,
\nonumber \\
&& \hspace{-100pt} + \,
\alpha \rho_{12} \left(
\frac{\partial^2}{\partial y \, \partial x_{12}} \, + \,
 \alpha \, \Box_y \right)
\left.\raisebox{18pt}{\hspace{-3pt}}\right\}
\, O_{2d_0 ,\hspace{1pt} \ell} \hspace{-1pt}
(y ;\hspace{1pt} x_{12})
\end{eqnarray}
by integrating by parts the term involving $\frac{\partial}{\partial {\alpha}}$.

\medskip

In the case $D=d=4$ ($d_0=1$) of interest
we find a 3-parameter family of solutions:
\begin{eqnarray}\label{new:2.29}
f_{1} \left( s,\, t \right) = && \hspace{-15pt}
a_0 \, i_0 \left( s,\, t \right) +
a_1 \, i_1 \left( s,\, t \right) +
a_2 \, i_2 \left( s,\, t \right)
\, , \quad \nonumber \\
i_0 \left( s,\, t \right) := && \hspace{-15pt}
1+t^{-1}
\, , \quad
i_1 \left( s,\, t \right) :=
\left( \frac{1-t}{t} \right)^2
\left( 1 + t - s \right) - 2 \, \frac{s}{t}
\, , \quad \raisebox{20pt}{} \nonumber \\
i_2 \left( s,\, t \right) := && \hspace{-15pt}
\frac{(1-t)^4}{t^3}
\left( 1+t-2s \right) \! - \! 6 s \frac{(1-t)^2}{t^2}
\! + \! \frac{s^2}{t^2} \left( 1+t \right) \!\left( 1 + \frac{(1-t)^2}{t} \right)
\! . \quad\quad\ \ \ \raisebox{20pt}{}
\end{eqnarray}
In both cases, $d=2$ and $d=4$ (\(D=4\)), we can compute $B_{1\ell} (d)$ from
(\ref{eq2.18}). The result is
\begin{equation}
\label{eq2.25}
B_{1\ell} (2) = \frac{2a_0}{\left( {4\ell \atop 2\ell} \right)} \quad
\end{equation}
(we have used $c$ for $a_0$ in \cite{NST1})
\begin{eqnarray}
\label{eq2.26}
B_{1\ell} (4)
\, = \,
\frac{1}{\left( {4\ell \atop 2\ell} \right)}
\left[\raisebox{11pt}{} 2 a_0 +
2\ell (2\ell + 1) \!
\left(\raisebox{10pt}{} 2 a_1 + (2\ell + 3)(\ell - 1) a_2 \right) \right] \, .
\end{eqnarray}

The condition for the absence of a $d=2$ scalar field in the OPE is given by
$a_0 = 0$ implying the vanishing of $V_1$ for $x_{12} \rightarrow 0$:
\begin{equation}
\label{eq2.24}
a_0 = 0 \Rightarrow V_1 (x_1 , x_2) = C \, x_{12}^{\mu} x_{12}^{\nu} T_{\mu\nu}
(x_1 , x_2) \, .
\end{equation}
The proportionality coefficient $C$ can be chosen in such a way that
$T_{\mu\nu} (x) := T_{\mu\nu} (x,x)$ is the stress energy tensor of the
theory, normalized by the standard Ward-Takahashi identity (cf.~\cite{MS}).

We note that the system (\ref{eq2.18}) is overdetermined: each $B_{1\ell}$
has to satisfy two conditions to fit the coefficients to $(1-t)^{2\ell}$ and
$(1-t)^{2\ell+1}$. Thus the existence of a solution provides a non-trivial
consistency check. The result (\ref{eq2.25}) was proven analytically using
the integral representation for the hypergeometric functions (see Appendix A
of \cite{NST1}); Eq.~(\ref{eq2.26}) was derived analytically for small values
of $\ell$ and verified numerically for $2\ell \leq 300$.

Wightman (i.e. Hilbert space) positivity implies
\begin{equation}
\label{eq2.27}
B_{\kappa \ell} = A_{\kappa 2 \ell} \, C_{\kappa 2 \ell} = N_{\kappa 2 \ell}
\, C_{\kappa 2 \ell}^2 \geq 0
\end{equation}
(since $C_{\kappa 2 \ell}$ is real, according to (\ref{eq2.12}), while
$N_{\kappa 2 \ell}$ is the normalization of the positive definite 2-point
function (\ref{eq2.6}) so it should be positive). The condition $B_{1\ell} >
0$ is indeed verified for $a_0>0$, $a_1 > 0$, $a_2 \geq 0$. This is only a
necessary condition for Wightman positivity. A necessary and sufficient
positivity condition was established in the $d=2$ case; it says that $c$
($= a_0$)
should be a positive integer (see \cite{NST1} Theorem~5.1). As the
ingredients in the derivation of this result have a more general significance
we review them in Appendix~\ref{apA} with an eye to a possible generalization to
the
theory of the Lagrangian field ${\cal L} (x)$ (of dimension $d=4$).

Knowing the singular part of the OPE (\ref{eq2.1}) allows, after {\it
crossing symmetrization} to reconstruct the complete 4-point function. The
result for the $d=2$ case is:
\begin{eqnarray}
\label{eq2.28}
&&\langle 1234 \rangle (= \langle 0 \mid \phi (x_1) \phi (x_2) \phi (x_3) \phi
(x_4) \mid 0 \rangle) \nonumber \\
&&= (1 + s_{23} + s_{13}) \left\{ \langle 12 \rangle \langle 34 \rangle +
\frac{1}{2} (12)(34) \langle 0 \mid V_1 (x_1 , x_2) V_1 (x_3 , x_4) \mid 0
\rangle \right\} \nonumber \\
&&= \langle 12 \rangle \langle 34 \rangle + \langle 13 \rangle \langle 24
\rangle + \langle 23 \rangle \langle 14 \rangle + w^t (x_1 , \ldots , x_4)
\end{eqnarray}
where, in the conventions of \cite{NST1} (for $d=2$),
\begin{equation}
\label{eq2.29}
\langle j\ell \rangle = \frac{c}{2} (j\ell)^2 \, , \ \left( (j\ell) =
\frac{1}{4\pi^2 \rho_{j\ell}} \, , \ j < \ell \right) \, ,
\end{equation}
\begin{eqnarray}
\label{eq2.30}
w^t (x_1 , x_2 , x_3 , x_4) &= &c \{ (12)(23)(34)(14) + (13) (23) (24) (14)
\nonumber \\
&+ &(12)(13)(24)(34) \}
\end{eqnarray}
for
\begin{equation}
\label{eq2.31}
\langle 0 \mid V_1 (x_1 , x_2) V_1 (x_3 , x_4) \mid 0 \rangle = c \{ (13)(24) +
(14)(23) \} \, .
\end{equation}

\medskip

\textit{Remark 2.2.}
The factor $\frac{1}{2}$ in the second term in the braces
in Eq. (\ref{eq2.28}) reflects a special case of the following
phenomenon. In general, we wish that the symmetrized
contribution of $V_{\kappa}(x_1,x_2)$ to the truncated 4-point function,
\begin{equation}\label{eq2.32}
F_{\kappa}(x_1,\, x_2,\, x_3,\, x_4) \, = \,
S \left\{ (12)^{d-\kappa}(34)^{d-\kappa}
\langle 0 \mid	V_{\kappa}(x_1,x_2)V_{\kappa}(x_3,x_4) \mid 0 \rangle \right\}
\, , \quad
\quad
\end{equation}
where $S$ stands for an yet unspecified symmetrization,
\(\kappa =1,\, \dots ,\, d-1\)
is not just symmetric under any permutation of its
arguments but that the difference
\[
F_{\kappa}(x_1,x_2,x_3,x_4)-(12)^{d-\kappa}(34)^{d-\kappa}
\langle 0 \mid	V_{\kappa}(x_1,x_2)V_{\kappa}(x_3,x_4)\mid 0 \rangle
\]
is of order $(12)^{d-\kappa}(34)^{d-\kappa}s$ (or smaller) for \(s\to 0\).
This second
condition forces us to use the standard symmetrization
\(\langle 12\rangle\langle 34\rangle\mapsto (1+s_{23}+s_{13})
\langle 12\rangle\langle 34\rangle=
\langle 12\rangle\langle 34\rangle+\langle 13\rangle\langle 24\rangle+
\langle 14\rangle\langle 23\rangle\)
for the disconnected term in the middle part of (\ref{eq2.28}) while applying
\(\frac{1}{2}\,\left( 1+s_{23}+s_{13} \right)\) to the second term.
We shall see in Sec.3.2
below that there exist a basis of rational harmonic functions whose
symmetrized contribution satisfying the above condition is
proportional to the standard symmetrization.
We further note that the function $F_1$, which is rational by itself,
can be viewed as providing a {\it minimal model} for the truncated
4-point function of $\phi$.

\section{General form of the 4-point function of ${\cal L} (x)$. Operator
content}\label{sec3}
\setcounter{equation}{0}

\subsection{%
	 s-channel contributions to the 4-point function for arbitrary twists}\label{ssec3.1}

We shall now exploit a result of Dolan and Osborn \cite{DO} which permits to
write down closed form expressions for $f_{\kappa}(s,t)$
(\ref{eq2.16}) given the rational
4-point function for $d=4(=D)$. More precisely, keeping in mind the
analysis of s-channel positivity we shall study the difference
\begin{eqnarray}
\label{eq3.1new}
&
w \left( x_1,\, x_2,\, x_3,\, x_4 \right) -
\langle 12 \rangle\langle 34 \rangle \, = \,
\langle 13 \rangle\langle 24 \rangle +
\langle 14 \rangle\langle 23 \rangle +
w^t \left( x_1,\, x_2,\, x_3,\, x_4 \right) \, = &
\nonumber \\ & = \,
\left( 2\pi \right)^{-8}
\left( \rho_{12}\rho_{23}\rho_{34}\rho_{14} \right)^{-2}
\left[ \frac{\textstyle 1}{\textstyle st}
P \left( s,\, t \right) + N^2 s^2 \left( t^2 +t^{-2} \right) \right]
. & \hspace{-10pt}
\end{eqnarray}
with \(P(s,t)(=P_4(s,t))\), a polynomial in s and t of overall degree 5
satisfying the crossing symmetry conditions (\ref{eq1.8})(\ref{eq1.9})
(see (\ref{eq3.28new}) below).
The rational
function in the brackets in the right hand side of (\ref{eq3.1new}) can be expressed
as an infinite sum of even twist contributions
\begin{equation}
\label{eq3.2new}
N^2 \, s^3 \, t \left( t^2 + t^{-2} \right) \, + \,
P \left( s,\, t \right) \, = \,
t^3 \mathop{\sum}\limits_{\kappa \, = \, 1}^{\infty} \,
s^{\kappa-1} \, f_{\kappa} \left( s,\, t \right)
\, , \qquad
\end{equation}
thus extrapolating (\ref{eq2.16}) beyond the range
\(\kappa=1,\, 2,\, 3\).
Albeit $f_{\kappa}$ are, in general, not rational functions of s and t they are
determined by the polynomial $P$ and the constant $N^2$.
Indeed Eq. (3.10) of \cite{DO} allows to write down the
following extension of (\ref{eq2.22new}) (\(d_0=1\)) to any
\(\kappa>1\):
\begin{eqnarray}
\label{eq3.3new}
f_{\kappa} \left( s,\, t \right) \, = \, \frac{1}{u-v} \,
\left\{\raisebox{11pt}{}\right. && \hspace{-20pt}
F \left( \kappa-1,\, \kappa-1;\, 2\kappa -2;\, v  \right)
g_{\kappa} \left( u \right)
\, -
\nonumber \\ && \hspace{-20pt} - \,
F \left( \kappa-1,\, \kappa-1;\, 2\kappa -2;\, u  \right)
g_{\kappa} \left( v \right)
\hspace{-2pt}\left.\raisebox{11pt}{}\right\}
\, . \qquad
\end{eqnarray}
We note that the normalization $N$ contributes to \(\kappa\geq d \ (=4)\) only. Then it
has to be taken into account when verifying Wightman positivity condition
(\ref{eq2.27}).
The hypergeometric functions $F(a,a;2a;v)$, \(a=1,\, 2,\, \dots\), are expressed
as linear combinations of $\log(1-v)$ with rational in $v$ coefficients:
\begin{eqnarray}
F \left( 1,\, 1;\, 2;\, v \right) \, = && \hspace{-15pt}
\mathop{\sum}\limits_{n \, = \, 1}^{\infty} \frac{v^{n-1}}{n} \, = \,
\frac{1}{v} \, \ln \frac{1}{1-v}
\, , \qquad
\nonumber \\
F \left( 1,\, 1;\, 2;\, v \right) \, = && \hspace{-15pt}
\mathop{\sum}\limits_{n \, = \, 1}^{\infty}
\frac{6\, n\, v^{n-1}}{\left( n+1 \right)\left( n+2 \right)} \, = \,
\frac{6}{v^2} \, \left( \frac{2-v}{v} \ln \frac{1}{1-v} -2 \right)
\, , \quad etc.
\qquad
\nonumber
\end{eqnarray}
The OPE of the bilocal field $V_{\kappa}(x_1,x_2)$ in terms of twist $2\kappa$ rank $2\ell$
symmetric traceless tensors corresponds, according to (\ref{eq2.18}), to an
expansion of ${g}_{\kappa}(u)$ in terms of hypergeometric functions of the same
type:
\begin{equation}
\label{eq3.4new}
g_{\kappa} \left( u \right)
\, = \, u f_{\kappa} \left( 0,\, 1-u \right) \, = \,
\mathop{\sum}\limits_{\ell \, = \, 0}^{\infty}
\, B_{\kappa\ell} \, u^{2\,\ell+1} \,
F \left( 2\ell+\kappa,\, 2\ell+\kappa;\, 4\ell+\kappa;\, u \right)
\,
\end{equation}
(Eq. (\ref{eq3.4new}) is obtained from (\ref{eq3.3new}) just using
${g}_{\kappa}(0)=0$, \(F(a,a;2a;0)=1\).)
The functions $f_{\kappa}(0,t)$ can be determined
recursively from the relations
\begin{eqnarray}
\label{eq3.5new}
f_{\kappa} \left( 0,\, t \right) \, = && \hspace{-15pt}
\mathop{\lim}\limits_{s \,\to\, 0} \,
\left\{\raisebox{15pt}{\hspace{-2pt}}\right.
s^{1-\kappa} \left[\raisebox{15pt}{\hspace{-2pt}}\right.
t^{-3} \, P \left( s,\, t \right)
- \sum_{\nu \, = \, 1}^{\kappa-1}
s^{\nu-1} \, f_{\nu} \left( s,\, t \right)
\left.\raisebox{15pt}{\hspace{-2pt}}\right]
\left.\raisebox{15pt}{\hspace{-2pt}}\right\}
\, , \quad
\nonumber \\
f_1 \left( 0,\, t \right) \, = && \hspace{-15pt}
t^{-3} \, P \left( 0,\, t \right)
\, .
\end{eqnarray}
and (\ref{eq3.3new}) which imply that the operator $\widehat{h}$,
\begin{equation}
\label{eq3.6new}
\widehat{h} \, t^{-3} \, P \left( s,\, t \right) \! := \!
\frac{1}{u \! - \! v} \!
\left\{\raisebox{15pt}{\hspace{-3pt}}\right.
\frac{u}{\left( 1 \! - \! u \right)^3}
\, P \left( 0,\, 1 \! - \! u \right)
\! - \! \frac{v}{\left( 1 \! - \! v \right)^3}
\, P \left( 0,\, 1 \! - \! v \right)
\left.\raisebox{15pt}{\hspace{-3pt}}\right\}
\! = \! f_{1} \left( s,\, t \right)
\end{equation}
defines a (rational) {\it harmonic projection} of the product
$t^{-3}P(s,t)$. In fact, the function $(13)(24)f_1(s,t)$ is harmonic in both
$x_{12}$ and $x_{34}$ for fixed $x_{23}$ while its difference with
$(13)(24)t^{-3}P(s,t)$
vanishes on either light cone (${\rho}_{12}=0$ and ${\rho}_{34}=0$) since
$t^{-3}P(0,t)=f_1(0,t)$ according to (\ref{eq3.5new}).
Thus Eq. (\ref{eq3.3new}) can be viewed as an
extension of the notion of harmonic projection to arbitrary twists.

\subsection{A basis of crossing symmetrized conformal harmonic
functions}\label{ssec3.2new}

We shall now display a basis ${j_{\nu}(s,t)}$ of rational solutions of the
{\it conformal Laplace equation} (\ref{new2.21}),
\begin{equation}
\label{eq3.7new}
\Delta \, j_{\nu} \left( s,\, t \right) \, = \, 0
\, , \ \, \nu\, = \, 0,\, 1,\, 2
\ \,
(
\Delta = s \frac{\partial^2}{\partial s^2} + t
\frac{\partial^2}{\partial t^2} +
(s+t-1) \frac{\partial^2}{\partial s \partial t} +
2 \, \frac{\partial}{\partial s} + 2\,\frac{\partial}{\partial t}
) \, .
\end{equation}
that are symmetric under $s_{12}$
\begin{equation}
\label{eq3.8new}
t^{-1} \, j_{\nu} \left( \frac{s}{t} ,\,  \frac{1}{t} \right) \, = \,
j_{\nu} \left( s,\, t \right)
\, \qquad
\end{equation}
and are eigenfunctions of the operator
\begin{equation}
\label{eq3.9new}
\widehat{h} \, \left( 1+s_{23}+s_{13} \right)
\, . \qquad
\end{equation}
We shall verify that such a basis is given by
\begin{eqnarray}
\label{eq3.10new}
j_{\alpha} \left( s,\, t \right) \, = && \hspace{-15pt}
i_{\alpha} \left( s,\, t \right)
\, , \quad  \alpha \, = \, 0,\, 1 \, , \quad
j_2 \left( s,\, t \right) \, = \, i_1 \left( s,\, t \right)
+ i_2 \left( s,\, t \right)
\, , \quad
\nonumber \\
t^{3} \, j_2(s,t) \, = && \hspace{-15pt} (1+t^{3})(1-s-t)^{2}-5 \, s \, t \, (1-t)^2
\, , \
\end{eqnarray}
where $i_{\nu}$ are defined in (\ref{new:2.29}).
It will become clear on the way that
$i_2$ is not an eigenvector of the operator (\ref{eq3.9new}) as
$j_1$ and $j_2$ correspond to different eigenvalues. Set indeed
\begin{equation}
\label{eq3.11new}
I_{\nu} \, = \, \left( 1 + s_{23} + s_{13} \right)
\left\{ t^3 \, i_{\nu} \left( s,\, t \right) \right\}
\, , \quad
\nu \, = \, 0,\, 1,\, 2 \, \qquad
\end{equation}
we find
\begin{eqnarray}
\label{eq3.12new}
I_0 (s,t) = && \hspace{-15pt}
s^2 (1+s) + t^2 (1+t) + s^2 t^2 (s+t) \, , \nonumber \\
I_1 (s,t) =  && \hspace{-15pt}
s (1+s) (1-s)^2 + t (1+t) (1-t)^2 + \nonumber \\
&& \hspace{-15pt} + \, st [(s-t) (s^2-t^2) -2\mathrm{Q}_1]
\, , \quad \mathrm{Q}_1 \, := \, 1+s^2+t^2 \nonumber \\
I_2 (s,t) = && \hspace{-15pt}
(1+s) (1-s)^2 (2-3s+2s^2) + (1+t) (1-t)^2 (3-3t+2t^2) - 2 + \nonumber \\
&& \hspace{-15pt} + st \left[ 4\mathrm{Q}_1-
\left( s+t \right)\left( 5s^2 - 8st + 5t^2 \right) \right] \, .
\qquad
\end{eqnarray}
The simplest way to compute the harmonic projection
\(\widehat{h}\left( t^{-3}I_{\nu}(s,t) \right)\)
consists in looking at the ``initial conditions'' for $s=0$.
We see that while
\(I_{\alpha} \left( 0,\, t \right) \, = \, t^3 i_{\alpha} \left( 0,\, t \right)\)
for \(\alpha \, = \, 0,\, 1\,\), we have
\begin{equation}
\label{eq.3.13new}
I_{2}(0,t)=2\,(1-t)^{4}(1+t)+t(1-t)^{2}(1+t)=
t^{3}\left( 2\, i_2(0,t)+i_1(0,t) \right)
\, . \qquad
\end{equation}
It follows that $i_2$ is {\it not} an eigenfunction of the operator
(\ref{eq3.9new}) but $j_2$ (\ref{eq3.10new}) is one (with eigenvalue 2).
This suggests introducing
a new basis of symmetrized twist two polynomial factors in the truncated
4-point function:
\begin{eqnarray}
\label{eq3.14new}
J_{\alpha} (s,t) \, (= && \hspace{-15pt} I_{\alpha} \left( s,\, t \right) )
\, = \, \left( 1 + s_{23} + s_{13} \right)
\left( t^3 j_{\alpha} \left( s,\, t \right) \right) \, , \quad
\alpha \, = \, 0,\, 1,\,
\nonumber \\
J_2 (s,t) = && \hspace{-15pt}
\frac{1}{2} \left( 1 + s_{23} + s_{13} \right)
\left( t^3 j_2 \left( s,\, t \right) \right) \, = \,
\frac{1}{2} \left( I_1 \left( s,\, t \right) + I_2 \left( s,\, t \right) \right)
\, = \,
\nonumber \\
= && \hspace{-15pt}
(1-s)^{2}(1+s^{3})+(1-t)^{2}(1+t^3)-1+
\nonumber \\
&& \hspace{-15pt} + st \left[ \mathrm{Q}_1-
\left( s+t \right)\left( 2s^2 - 3st + 2t^2 \right) \right] \, ;
\qquad
\end{eqnarray}
the $J_{\nu}$ are distinguished by the properties
\begin{eqnarray}
\label{eq3.15new}
\widehat{h} \, \left\{ t^{-3} J_{\nu} (s,t) \right\}
\, = && \hspace{-15pt} j_{\nu} \left( s,\, t \right) \, , \quad
\nu \, = \, 0,\, 1,\, 2;\,
\nonumber \\
J_0 \left( s,\, t \right) - t^3 \, j_0 \left( s,\, t \right)
\, = && \hspace{-15pt}
s^2\, (1+t^3)+s^3\, (1+t^2)
\nonumber \\
J_1 \left( s,\, t \right) - t^3 \, j_1 \left( s,\, t \right)
\, = && \hspace{-15pt}
s\,(1-t)(1-t^3)-s^2\,(1+t^3)-s^3\,(1+t)^2+s^4\,(1+t)
\nonumber \\
J_2 \left( s,\, t \right) - t^3 \, j_2 \left( s,\, t \right)
\, =  && \hspace{-15pt}
s^3\,(1+t+t^2)-2s^4\,(1+t)+s^5
\, . \hspace{-50pt}
\end{eqnarray}
Comparing (\ref{eq3.15new}) with (\ref{eq3.5new})
we deduce that the differences \(J_1-t^3j_1\),
\(J_0-t^3j_0\), and \(J_2-t^3j_2\) involve twist 4 and higher,
twist 6 and higher, and twist 8 and higher, respectively.
The expressions (\ref{new:2.29}) for
$f_1$ and the corresponding crossing symmetric amplitude
$F_1$ can be rewritten in the basis (\ref{eq3.10new})
(\ref{eq3.14new}) with the result
\begin{eqnarray}
\label{eq3.16new}
f_1 \left( s,\, t \right)
\, = && \hspace{-15pt} a_0 \, j_0 \left( s,\, t \right) +
a_{12} \, j_1 \left( s,\, t \right) + a_2 \, j_2 \left( s,\, t \right)
\, , \quad a_{12} \, = \, a_1 - a_2 \, ,
\nonumber \\
F_1 \left( x_1,\, x_2,\, x_3,\, x_4 \right)
\, = && \hspace{-15pt}
\left( 12 \right)^2 \left( 23 \right)^2 \left( 34 \right)^2 \left( 14 \right)^2
\, \frac{1}{st} \, \times
\nonumber \\
&& \hspace{-15pt} \times
\left\{ a_0 \, J_1 \left( s,\, t \right) + a_{12} \, J_{12} \left( s,\, t \right)
+ a_2 \, J_2 \left(  s,\, t \right) \right\}
\, . \hspace{-50pt}
\end{eqnarray}
We note that the truncated 4-point function of the (vacuum) Maxwell
Lagrangian,
\begin{equation}
\label{eq3.9}
{\cal L}_0 (x) = - \frac{1}{4} : F_{\mu \nu} (x) F^{\mu \nu} (x) : \, ,
\end{equation}
(or a sum of mutually commuting expressions of this type) is a special case of
(\ref{eq3.16new}). Here $F_{\mu \nu}$ is a free electromagnetic field with 2-point
function
$$
\langle 0 \mid F_{\mu_1 \nu_1} (x_1) F_{\mu_2 \nu_2} (x_2) \mid 0 \rangle = 4
D_{\mu_1 \nu_1 \mu_2 \nu_2} (x_{12})
$$
\begin{equation}
\label{eq3.10}
= \{ \partial_{\mu_1} (\partial_{\mu_2} \eta_{\nu_1 \nu_2} - \partial_{\nu_2}
\eta_{\nu_1 \mu_2}) - \partial_{\nu_1} (\partial_{\mu_2} \eta_{\mu_1 \nu_2} -
\partial_{\nu_2} \eta_{\mu_1 \mu_2})\} (12) \, ;
\end{equation}
we have
\begin{equation}
\label{eq3.11}
4\pi^2 D_{\mu_1 \nu_1 \mu_2 \nu_2} (x) = R_{\mu_1 \mu_2} (x) R_{\nu_1 \nu_2}
(x) - R_{\mu_1 \nu_2} (x) R_{\nu_1 \mu_2} (x) \, ,
\end{equation}
the symmetric tensor $R(x)$ being defined in (\ref{eq2.8}). The OPE of the
product of two ${\cal L}_0$'s has the form
\begin{equation}
\label{eq3.12}
{\cal L}_0 (x_1) {\cal L}_0 (x_2) = \langle 12 \rangle_0 + \frac{8}{\pi^2
\rho_{12}^3} \, T (x_1 , x_2 ; x_{12}) + : {\cal L}_0 (x_1) {\cal L}_0 (x_2) :
\end{equation}
where $\langle 12 \rangle_0 = 3 (\pi \rho_{12})^{-4}$, and $T(x_1 , x_2 ; x_{12})$ (a
multiple of $V_1$) is the harmonic bilocal field
\begin{eqnarray}
\label{eq3.13}
T \left( x_1,\, x_2;\, x_{12} \right) \, = && \hspace{-10pt}
\frac{1}{4} : F^{\sigma \tau} (x_1) F_{\sigma \tau} (x_2) : x_{12}^2 -
x_{12}^{\mu} : F^{\sigma}_{\hspace{5pt}\mu} (x_1) F_{\sigma \nu} (x_2) : x_{12}^{\nu}
\, = \,
\nonumber \\
\, = && \hspace{-10pt}
\frac{1}{4} \, x_{12}^{\, 2} \left( r_{12} \right)^{\mu\nu}
\left( r_{12} \right)^{\sigma\tau}
: F_{\mu\sigma} (x_1) F_{\nu\tau} (x_2) :
\, .
\end{eqnarray}
In verifying \(\Box_1T(x_1,x_2;x_{12})=0\) one should use both the tracelessness of
$T$, i.~e. \(\Box_{y}T(x_1,x_2;y)=0\) and the two Maxwell equations,
\begin{equation}
\label{eq3.18new}
d \, F(x) \, = \, 0 \, = \, d \, ^* \hspace{-1pt}F(x) \, , \quad
\end{equation}
where $F(x)$ is the 2-form (\ref{eq1.12}).
The second expression for \(T \left( x_1,\, x_2;\, x_{12} \right)\) (\ref{eq3.13})
makes obvious its conformal invariance.
The 4-point function of ${\cal L}_0$ is computed from (\ref{eq3.9})--(\ref{eq3.11})
using
repeatedly the triple-product formula (of \cite{OP})
\begin{equation}
\label{eq3.14}
r(x_{12}) r(x_{23}) r(x_{13}) = r (X_{23}^1) \, , \qquad X_{23}^1 =
\frac{x_{13}}{\rho_{13}} - \frac{x_{12}}{\rho_{12}}
\end{equation}
(see Appendix B of \cite{NST1}). The contribution of $T$ to the 4-point function is
given by the solution (\ref{new:2.29}) of (\ref{new2.21}):
\begin{eqnarray}
\label{eq3.15}
&&\pi^4 \langle 0 \mid T (x_1 , x_2 ; x_{12}) \, T (x_3 , x_4 ; x_{34} ) \mid 0 \rangle
= (13)(24) f_{14} (s,t) \nonumber \\
&&\hbox{for} \ a_0 = 0 \, ,  \ a_1 = a_2 \, ;
\end{eqnarray}
it follows that $w_4^t$ from (\ref{eq3.1new}) assumes the form
\begin{eqnarray}
\label{eq3.16}
w_{{\cal L}_0}^t (x_1 , x_2 , x_3 , x_4) &= &(1+s_{12} + s_{23}) c_0 tr (D_{12}
D^{23} D_{34} D^{14}) \nonumber \\
&= &\frac{c_0}{8\,\pi^8} \, (\rho_{12} \rho_{23} \rho_{34} \rho_{14} )^{-2}
\left( st \right)^{-1} J_2 \left( s,\, t \right)
\end{eqnarray}
\begin{equation}
\label{eq3.17}
D_{12} := D_{\mu_1 \nu_1 \mu_2 \nu_2} (x_{12}) \, , \ D^{23} := D^{\mu_2 \nu_2 \mu_3
\nu_3} (x_{23})
\quad \mathrm{etc.} \quad
\end{equation}
corresponding to $a_1 = a_2 = 2^5 c_0$, $a_0 = a_{12} = 0$. In fact, a
finite sum of expressions of type (\ref{eq3.9}) will give rise to a discrete
subset of such values with a positive integer $c_0$.

\subsection{General form of $w^{t}_{4}$.
	Higher twist contributions}\label{ssec3.2}

The most general GCI truncated 4-point function of a \(d=4\) scalar field
contains, in addition to \(F_1\) (\ref{eq3.16new}),
two more terms corresponding to
crossing symmetrized twist 4 contributions. They can be written as
$s \, t \, Q_i(s,t)$, \(i=1,\, 2\), where $Q_i$ are crossing symmetric polynomials
of degree 2:
$$
Q_1 = 1 + s^2 + t^2 \, , \ Q_2 = s+t+st \, ,
$$
\begin{equation}
\label{eq3.4}
t^2 Q_j \left( \frac{s}{t} , \frac{1}{t} \right) = Q_j (s,t) = s^2 Q_j \left(
\frac{1}{s} , \frac{t}{s} \right) \, , \ j =1,2 \, .
\end{equation}
Thus, the polynomial $P(s,t)$ of Eq.~(\ref{eq3.1new}) can be presented in the form:
\begin{equation}
\label{eq3.28new}
P(s,t) \, = \, a_{0}J_{0}(s,t)+a_{12}J_{1}(s,t)+a_{2}J_{2}(s,t)
+s \, t \left[ b_{1}Q_{1}(s,t)+b_{2}Q_{2}(s,t) \right] \, . \
\end{equation}

\medskip

\textit{Remark 3.1.}
The above basis of crossing symmetric polynomials ${J_{\nu},Q_j}$
and the ${P_{\nu},Q_j}$ basis of \cite{NST1}, where
$$
P_0 (s,t) = 1 + s^5 + t^5 \, , \ P_1 (s,t) = s(1+s^3) + t(1+t^3) + st (s^3 +
t^3) \, ,
$$
\begin{equation}
\label{eq3.5}
P_2 (s,t) = s^2 (1+s) + t^2 (1+t) + s^2 t^2 (s+t) \ (= I_0 (s,t)) \, ,
\end{equation}
are related by:
\begin{equation}
\label{new3.30}
J_0=P_2\, , \quad
J_1=P_1-P_2-2 \, s\, t \, Q_1 \, , \quad
J_2=P_0-2P_1+P_2+s\, t\, Q_1 \, . \
\end{equation}
We shall now apply the procedure outlined in Sec.3.1 to compute the twist
4 and 6 contributions in (\ref{eq3.28new})). It follows from (\ref{eq3.4new})
(\ref{eq3.15new}) and (\ref{eq3.28new})
that
\begin{eqnarray}\label{new3.30add}
f_2 \left( 0,\, 1\! -\! u \right) = && \hspace{-18pt}
a_{12}\hspace{-1pt} \left(\raisebox{14pt}{\hspace{-4pt}}\right.
1 \! - \! u \! +\frac{1}{\left( 1 \! - \! u \right)^3}
\left.\raisebox{14pt}{\hspace{-4pt}}\right) \! +
\left( b_1 \! - \! a_{12} \right)\hspace{-1pt}
\left(\raisebox{14pt}{\hspace{-4pt}}\right.
1\! +\frac{1}{\left( 1 \! - \! u \right)^2}
\left.\raisebox{14pt}{\hspace{-4pt}}\right) \! +
b_2\hspace{2pt} \frac{1}{1 \! - \! u}
\equiv
\nonumber \\
\equiv\frac{g_2 \left( u \right)}{u} = && \hspace{-15pt}
\mathop{\raisebox{0pt}{\large $\sum$}}\limits_{\ell \, = \, 0}^{\infty}
B_{2\hspace{1pt}\ell} \
u^{2\ell} \
F \left( 2\ell+2,\, 2\ell+2;\, 4\ell+4;\, u \right)
\, \qquad
\end{eqnarray}
yielding
\begin{equation}\label{new3.31}
B_{2\hspace{1pt}\ell} =
{4\ell +1 \choose 2\ell }^{-1}
    \left[\raisebox{12pt}{\hspace{1pt}}
		\ell \left( \ell +1 \right)\left( 2\ell +1 \right)
		\left( 2\ell +3 \right) a_{12} +
		2\left( \ell +1 \right)\left( 2\ell +1 \right) b_1 + b_2 \right].
\end{equation}
We see, in particular,
that the absence of a scalar field of dimension 4 in the OPE of
two ${\cal L}$'s implies
\begin{equation}\label{3.33}
B_{20} = 2 \, b_{1}+b_{2} = 0 \quad ( \Rightarrow \langle 123 \rangle = 0)
\, . \
\end{equation}
The difference between $P(s,t)$ (\ref{eq3.28new}) and the twist 2
(Eq.~(\ref{eq3.16new})) and
twist 4 part, computed from (\ref{eq3.3new}),
defines $t^{3}f_{3}(0,t)$ as the coefficient to the $s^2$ term:
\begin{equation}\label{eq3.32new}
P(s,t)-t^{3}[f_1(s,t)+sf_{2}(s,t)]=s^{2}t^{3}f_{3}(0,t)+O(s^3) \, . \
\end{equation}
A computer aided calculation (using Maple) gives
\begin{eqnarray}\label{eq3.34new}
\frac{\textstyle g_3 \left( u \right)}{\textstyle u}
\, = && \hspace{-15pt}
\mathop{\raisebox{0pt}{\large $\sum$}}\limits_{\ell \, = \, 0}^{\infty} \
B_{3\hspace{1pt}\ell} \
u^{2\ell} \
F \left( 2\ell+3,\, 2\ell+3;\, 4\ell+6;\, u \right)
\, = \,
\nonumber \\ = && \hspace{-15pt}
\frac{1}{2}\ a_{12}+a_0
\, + \, \frac{1}{2}\ \frac {b_2-b_1}{1-u}
\, + \, \frac{1}{2}\ \frac{b_1+2\,b_2}{\left( 1-u \right)^{2}}
\, + \, \frac{1}{2}\ \frac{a_{12}+2\,a_0}{ \left( 1-u \right)^{3}}
\, - \, \nonumber \\ && \hspace{-15pt}
\, - \, \frac{1}{2}\ \left( 2\,b_1 + b_2 \right)
\left(  \frac{1}{u} + \frac{2}{{u}^{2}} \right)
\, - \, \left( 2\,b_1+b_2 \right) \, \frac{\ln \left( 1-u \right)}{{u}^{3}}
\, , \qquad
\raisebox{17pt}{}
\end{eqnarray}
with
\begin{eqnarray}\label{eq3.35new}
B_{3\hspace{1pt}\ell} & =
\raisebox{0pt}{\Large \(\frac{1}{2} {4\ell +3 \choose 2\ell +1}\)}^{-1}
   \left[\raisebox{12pt}{\hspace{-10pt}}\right. &
	 \left( \ell +1 \right)\left( \ell +2 \right)
	 \left( 2\ell +1 \right)\left( 2\ell +3 \right)\left( 2a_0+a_{12} \right)+
\nonumber \\
& & +
   2\left( \ell +1 \right)\left( 2\ell +3 \right)\left( b_1+2 b_2 \right)
	 - 2 b_1 - b_2 \left.\raisebox{12pt}{\hspace{-2pt}}\right].
\end{eqnarray}
Remarkably, $g_3 \left( u \right)$ is rational precisely when the condition
(\ref{3.33}), reflecting the electric-magnetic duality, is satisfied.

\section{Implications for the gauge field Lagrangian}\label{sec4new}
\setcounter{equation}{0}

\subsection{Restrictions from OPE, Hodge duality,
	and Wightman positivity}\label{sec4.1new}

As discussed in the Introduction the gauge field Lagrangian (\ref{eq1.11}) is
characterized by the absence of scalar fields of dimension 2 and 4 in the OPE
of two ${\cal L}$'s. According to (\ref{eq2.24}) and (\ref{3.33})
this implies
\begin{equation}\label{eq4.1new}
a_0=0=2b_1+b_2.
\end{equation}
This allows to write the polynomial $P(s,t)$ in (\ref{eq3.1new}) as
\begin{equation}\label{eq4.2new}
P(s,t)=a'J_1(s,t)+aJ_2(s,t)+b[Q_1(s,t)-2Q_2(s,t)]
\end{equation}
where we have set $a_2=a$, $a_{12}=a'$,
\(b_1=b(=-\frac{\textstyle b_{2}}{\textstyle 2})\); the difference
\begin{equation}\label{eq4.3new}
Q_1(s,t)-2Q_2(s,t)=(1-s-t)^{2}-4\, s\, t
\end{equation}
is characterized by having a (second order) zero at $t=1$ for $s=0$
(and being negative for euclidean $x_{ij}$).

The restrictions on the three remaining constants, $a$, $a'$, and $b$, coming
from the positivity condition (2.33) for $B_{k l}$ given by (\ref{eq2.26})
(\ref{new3.31}) and (\ref{eq3.35new}) are
\begin{equation}\label{eq4.4new}
a>0 , \ \, a' \geq 0;
\end{equation}
\((\ell +1)(2\ell +1) \, a'+2b \geq 0\), for \(l >0\);
\((\ell +2)(2\ell +1) \, a' \geq 6b\)
for \(\ell \geq 0\). The last two inequalities
imply
\begin{equation}\label{eq4.5new}
- 3 \, a' \, \leq \, b \, \leq \, \frac{1}{3} \ a'
\, . \quad
\end{equation}
In particular, if \(a'=0\) it would follow that also \(b=0\) and we would end up
with the truncated 4-point function that is a multiple of the one of the
free electromagnetic Lagrangian (\ref{eq3.9}).

\medskip

\textit{Remark 4.1.}
If we allow for a positive $a_0$ in the 4-point function
(including, say, a scalar field contribution in the Lagrangian) then,
according to (\ref{eq3.35new}), the restriction
(\ref{eq4.5new}) would be replaced by a more
general one
\begin{equation}\label{eq4.6new}
-6a' \leq b \leq 3(a_0+\frac{a'}{2})
\end{equation}
which leaves room for a (non-zero) positive $b$ even for \(a'=0\).
A numerical analysis indicates that the most general GCI 4-point function
involves a non-trivial open domain in the 5-dimensional projective space of the
parameters $a_{\nu}$, $b_{j}$ and the 2-point function normalization, $N^2$,
defined up to a common positive factor, in which Wightman positivity is
verified for all twists.

\subsection{Difficulties in exploiting the compositeness
	of ${\cal L}$}\label{ssec4.2new}

It is instructive to understand the restrictions, coming from (\ref{eq4.1new}),
within an axiomatic treatment of non-abelian gauge field theory. At the same
time we shall try to answer the question: can we extract more information from
the expression (\ref{eq1.11}) for ${\cal L}$ in terms of $F_{\mu\nu}$ than
just saying that the OPE of ${\cal L} (x_1) {\cal L} (x_2)$ contains neither
${\cal L}$ nor a scalar field of dimension 2?

In (perturbative) quantum electrodynamics amplitudes with an odd number of
photon external legs (and no external charged particles) vanish because of
charge conjugation invariance (Furry's theorem). The general conformally
invariant 3-point function of a Maxwell field $F_{\mu\nu} (x)$, on the other
hand, violates local commutativity of Bose fields and should hence be also set
equal to zero (without having to assume any discrete symmetry). This last
argument fails for a non-abelian gauge field
\begin{equation}
\label{eq3.21}
F(x,\omega) := \frac{1}{2} \omega^{\mu \nu} F_{\mu\nu}^a (x) t_a \, , \ [t_a ,
t_b] = if_{abc} \, t_c \, , \ (\omega^{\mu\nu} = - \omega^{\nu\mu})
\end{equation}
where $t_a$ are orthonormal hermitian matrices generating the defining
representation of a (compact, semi-simple) non-abelian gauge group $G$,
$f_{abc}$ is the totally antisymmetric tensor of (real) structure constants of
the Lie algebra ${\cal G}$ of $G$ ($f_{abc} = \varepsilon_{abc}$, the
Levi-Civita tensor for ${\cal G} = su(2)$, $t_a = \frac{1}{2} \, \sigma_a$,
$a,b,c = 1,2,3$), $\omega$ is a constant skew-symmetric tensor (or
differential form) introduced for notational convenience. While the general
(gauge and) conformally invariant 2-point function of $F^a (x,\omega) =
\frac{1}{2} \, \omega^{\mu\nu} F_{\mu\nu}^a (x)$ coincides with the free
Maxwell one, (\ref{eq3.10}),
\begin{equation}
\label{eq3.22}
\langle 0 \mid F^a (x_1 , \omega_1) F^b (x_2 , \omega_2) \mid 0 \rangle = N_F
\delta^{ab} D (x_{12} ; \omega_1 , \omega_2)
\end{equation}
where we have introduced a contracted form of (\ref{eq3.11}),
\begin{equation}
\label{eq3.23}
4\pi^2 D (x ; \omega_1 , \omega_2) = R_{\mu_1 \mu_2} (x) R_{\nu_1 \nu_2} (x)
\omega_1^{\mu_1 \nu_1} \omega_2^{\mu_2 \nu_2} \, ,
\end{equation}
 there is a 2-parameter family of local gauge and conformally invariant 3-point
functions:
\begin{equation}
\label{eq3.24}
W^{abc} (x_1 , \omega_1 , x_2 , \omega_2 , x_3 , \omega_3) = f^{abc} (N_1
W^{(1)} + N_2 W^{(2)}) \, ,
\end{equation}
\begin{eqnarray}
\label{eq3.25}
W^{(1)} (x_1 , \omega_1 , x_2 , \omega_2 , x_3 , \omega_3) &= &(X_{23\mu_1}^1
R_{\nu_1 \mu_3} (x_{13}) X_{31 \mu_2}^2 R_{\nu_2 \nu_3} (x_{23}) \nonumber \\
&- &X_{23 \mu_1}^1 R_{\nu_1 \mu_2} (x_{12}) X_{12 \mu_3}^3 R_{\nu_2 \nu_3}
(x_{23}) \nonumber \\
&+ &X_{31 \mu_2}^2 R_{\mu_1 \nu_2} (x_{12}) X_{12 \mu_3}^3 R_{\nu_1 \nu_3}
(x_{13})) \nonumber \\
&&\omega_1^{\mu_1 \nu_1} \omega_2^{\mu_2 \nu_2} \omega_3^{\mu_3 \nu_3} \, ,
\end{eqnarray}
\begin{eqnarray}
\label{eq3.26}
W^{(2)} (x_1 , \omega_1 , x_2 , \omega_2 , x_3 , \omega_3) &= &R_{\mu_1 \mu_2}
(x_{12}) R_{\nu_1 \mu_3} (x_{13}) R_{\nu_2 \nu_3} (x_{23}) \nonumber \\
&&\omega_1^{\mu_1 \nu_1} \omega_2^{\mu_2 \nu_2} \omega_3^{\mu_3 \nu_3} \, ,
\end{eqnarray}
where $R_{\mu\nu}$ and $X_{12}^3$ are defined in (\ref{eq2.8}) and
(\ref{eq2.9}).

The expression (\ref{eq3.24}) can, sure, be used to produce a non-zero 3-point
function of ${\cal L}$ (\ref{eq1.11}). Thus, the second condition (\ref{eq4.1new}) is
not, at first sight, an automatic consequence of (\ref{eq1.11}), conformal
invariance and locality, Hodge duality providing an independent restriction on
correlation functions. It turns out, however, that {\it the combination of
(\ref{eq3.22}) and (\ref{eq3.24})
implies the existence of a field ${\cal I}$ in the OPE of
\(F (x_1) \otimes F (x_2)\) transforming under an
indecomposable representation of ${\cal C}$ (which would severely complicate
the use of conformal invariance).}

This statement follows from the observation that the 3-point function
(\ref{eq3.24}) does not satisfy the free Maxwell equation $d^* F = 0$
for any non-zero choice of $N_1$ and $N_2$
while the $2$--point function does:
$$
\langle 0 \mid d^* F (x_1) \otimes F (x_2) \mid 0 \rangle = 0 ,\,
$$
\begin{equation}
\label{eq3.27}
\langle 0 \mid d^* F (x_1) \otimes F (x_2)\otimes F (x_3) \mid 0 \rangle \neq 0
\, .
\end{equation}
It follows that the OPE of $F
(x_2)\otimes F (x_3)$ should involve a local field ${\cal I}$ that is not orthogonal
to $d^* F$. ${\cal I}$ cannot be a
derivative of $F$ because of the first equation (\ref{eq3.27}). On the other
hand, ${\cal I}$ cannot be an elementary conformal field transforming under an
inequivalent representation of ${\cal C}$ since then it should again be
orthogonal to $F$, and hence to $d^*F$ in accord with (\ref{eq3.27}).

Thus the vanishing of odd point correlation functions of ${\cal L}$ is a
natural property of the Lagrangian (\ref{eq1.11}) if the local observable
algebra is spanned by elementary conformal fields (and their derivatives).

\medskip

{\it Remark 4.2.} Let $A_{\mu}^a (x)$, $a=1,2,3$ be three commuting purely
longitudinal gauge potentials, -- i.e., generalized free fields such that
$$
\langle 0 \mid A_{\mu}^a (x_1) A_{\nu}^b (x_2) \mid 0 \rangle = \frac{1}{2} \,
\delta^{ab} r_{\mu\nu} (x_{12}) (12) = \frac{\delta^{ab}}{8\pi^2} \, R_{\mu\nu}
(x_{12}) \, ,
$$
\begin{equation}
\label{eq3.28}
\partial_{\mu} A_{\nu}^a (x) = \partial_{\nu} A_{\mu}^a (x) \, .
\end{equation}
Then both the 2-point function (\ref{eq3.22}) and the 3-point function
(\ref{eq3.26}) are reproduced by the corresponding $su(2)$ Yang-Mills curvature
tensor
\begin{eqnarray}
\label{eq3.29}
F_{\mu\nu}^a (x) &= &\partial_{\mu} A_{\nu}^a (x) - \partial_{\nu} A_{\mu}^a (x)
- g \varepsilon^{abc} : A_{\mu}^b (x) A_{\nu}^c (x) : \nonumber \\
&= &-g \varepsilon^{abc} : A_{\mu}^b (x) A_{\nu}^c (x) : \, .
\end{eqnarray}
This example is already excluded, however, by our requirement that no $d=2$
scalar field appears in the OPE of two ${\cal L}$'s (neither does $F$
(\ref{eq3.29}) satisfy the Yang-Mills equation with connection $A$). Even for
the more general $d=4$ composite field
\begin{equation}
\label{eq3.30}
{\cal L}_{\xi \eta} (x) = \xi : (A_{\mu}^a (x) A_a^{\mu} (x))^2 : - \eta :
A_{\mu}^a (x) A_b^{\mu} (x) A_{\nu}^b (x) A_a^{\nu} (x):
\end{equation}
(which includes (\ref{eq1.11}) with $F$ given by (\ref{eq3.29}) for $\xi = \eta
= \frac{g^2}{4}$) we find that the leading term in the OPE of two ${\cal L}$'s,
\begin{equation}
\label{eq3.31}
{\cal L}_{\xi\eta} (x_1) {\cal L}_{\xi\eta} (x_2) \approx 4(14 \xi^2 - 16
\xi\eta + 11 \eta^2) (12)^3 (A_{\mu}^a (x_1) r^{\mu\nu} (x_{12}) A_{\nu}^a
(x_2))
,
\end{equation}
involves a scalar field of dimension $d=2$ (with the same coefficient as the
2-point function of ${\cal L}_{\xi\eta}$ --  which is non-zero for any not
simultaneously zero real $\xi , \eta$).

\subsection{Concluding remarks}\label{sec4}

We have presented in the preceding sections two intertwined developments:
(i)~a~systematic study of the theory of a GCI local scalar field $\psi$ of any
(integer) dimension $d$ in terms of bilocal fields $V_{\kappa} (x_1 , x_2)$ of
dimension $(\kappa , \kappa)$ appearing in the OPE $\psi^* (x_1) \psi (x_2)$
(\ref{eq2.1}) (\ref{eq2.5new}); (ii) first steps in an attempt to construct in a
non-perturbative, axiomatic approach a conformally invariant fixed point of
a gauge field theory, formulated entirely in terms of gauge invariant
local observables of dimension 4: the Lagrangian density ${\cal L} (x)$ and the
stress-energy tensor $T_{\mu\nu} (x)$ or, rather, its polarized bilocal
counterpart that determines $V_1 (x_1 , x_2)$ according to (\ref{eq2.24}).

\medskip

(i) The use of bilocal fields simplifies substantially the analysis of the
operator content of GCI correlation functions. In particular, $V_1 (x_1 ,
x_2)$, defined as the (infinite) sum of twist 2 conserved symmetric tensor
fields, satisfies as a consequence the d'Alembert equation (\ref{eq2.20}).
This allows to compute the contribution of the correlator $\langle 0 \mid
V_1 (x_1 , x_2) V_1 (x_3 , x_4) \mid 0 \rangle$ to the (truncated) 4-point
function $w_4^t$ of ${\cal L}$. (Its expression for the $d=2$
neutral scalar field $\phi$ has been computed in \cite{NST1}.)
A {\it minimal model} for the connected part of the 4-point function of a
general neutral scalar field of dimension 4 is given by $F_1$
(\ref{eq3.16new}) which
is determined by its harmonic projection. A general procedure is outlined -
based on the Dolan-Osborn formula (\ref{eq3.3new}) - for computing the expectation
values (\ref{eq2.16}) of $V_{\kappa}$.
The case $\kappa =1$ is distinguished by the fact that the
conformal harmonic function $f_1(s,t)$ (and hence, the corresponding
symmetrized contribution $F_1$) is rational. For $\kappa >1$ $f_{\kappa}(s,t)$
are linear
combinations (with rational function coefficients) of
\begin{equation}\label{eq4.14new}
\log(1-u) \ \, \hbox{and} \ \, \log(1-v) \ \, \hbox{for} \ \, s=uv, \ \,
s+1-t=u+v \, . \
\end{equation}
We have displayed these functions (taking into account also the
additional term \(b_{1}Q_{1}+b_{2}Q_2\) in (\ref{eq3.28new})) and the associated
(infinite) OPE expansions in terms of symmetric traceless tensor fields of
twist $2\kappa$ for $\kappa=1,\, 2,\, 3$
(see Eqs.~(\ref{eq2.26}), (\ref{new3.30add})(\ref{new3.31}),
(\ref{eq3.34new})(\ref{eq3.35new}),
respectively). The same procedure applies to higher $\kappa$ as well, when $f_{\kappa}$
also depend on the sum of products of 2-point functions
\(\langle 13\rangle \langle 24\rangle +\langle 14\rangle \langle 23\rangle\)
(which introduces one more paprameter, the normalization $N$ of the
2-point function). The resulting expressions (together with other computer
aided results - concerning 6-point functions) will be presented in a
forthcoming publication, in collaboration with K.-H. Rehren.

\medskip

(ii) Conditions (\ref{sec4.1new}) exclude contributions of scalar fields of
dimension 2 and 4 in the OPE of \({\cal L}(x_1){\cal L}(x_2)\)
leaving us with a 3-parameter
family of truncated 4-point functions. We assert that ${\cal L}(x)$ then has
the properties of the Lagrangian density of a gauge field curvature
(without matter fields). The study of Wightman positivity for twists 4
and 6 contributions leads to a rather strong constraint (\ref{eq4.5new}) for the
remaining parameters. Should, in particular, the analysis of the 6-point
function of ${\cal L}$ yield the constraint \(a'=0\)
Eq.(\ref{eq4.5new}) would also imply \(b=0\)
and leave us with a multiple of the 4-point function of the Lagrangian of
a free abelian gauge field. This would confirm the general belief (see,
e.g. \cite{F} \cite{W03}) that a (pure) non-abelian gauge
theory necessarily involves a
mass gap, thus violating conformal invariance. By contrast, if we do not
impose (\ref{sec4.1new})~--~i.e., if we allow for the presence of
(at least) a scalar
field in the Lagrangian, a full account of Wightman positivity for the
4-point function appears to allow for an open set of the 5-dimensional
(projective) parameter space.

The evidence that we are displaying a gauge invariant 4-point function in a
(non-abelian) gauge theory is rather indirect. One verifies on a case by case
basis that any other renormalizable Lagrangian would involve fields of $d<4$
in the OPE and that is excluded by condition (\ref{eq4.1new}). The
difficulty in identifying the theory in terms of basic (gauge dependent) fields
like $F_{\mu\nu}^a$ (briefly reviewed in Sec.~\ref{ssec4.2new}) lies in the fact
that the model we are trying to construct is necessarily non-perturbative (if
it exists at all).
From this point of view our model is not incompatible with the $N=4$
supersymmetric Yang-Mills theory (see e.g. \cite{AEPS},
\cite{BKS}, \cite{BKRS}, \cite{EPSS} and references therein)
for sufficiently large value of the coupling constant
$g$, such that the anomalous dimension of the Konishi field
(which appears also in the OPE of two fields of the supermultiplet of the
stress-energy tensor) is a positive even integer.

\bigskip

\noindent {\bf Acknowledgments.} The authors thank Dirk Kreimer for a
stimulating discussion and Karl-Henning Rehren for an enlightening correspondence. N.N.
and I.T. acknowledge the hospitality of the Erwin
Schr\"odinger International Institute for Mathematical Physics (ESI) as well as
partial support by the Bulgarian National Council for Scientific Research under
contract F-828. The research of Ya.S. was supported in part by I.N.F.N., by the
EC contracts HPRN-CT-2000-00122 and -00148, by the INTAS contract 99-0-590 and
by the MURST-COFIN contract 2001-025492.
All three authors acknowledge partial support by the
Research Training Network within the Framework Programme 5 of the European
Commission under contract HPRN-CT-2002-00325 and
by a NATO linkage grant PST.CLG.978785.
I.T. thanks l'Institut des Hautes
Etudes Scientifiques (Bures-sur-Yvette), the Theory Division of CERN and
Section de Math\'{e}matiques, Universit\'{e} de Gen\`{e}ve for hospitality during
the final stage of this work.

\newpage

\appendix
\section{$\!\!\!\!\!\!\!$ppendix A. Vacuum representation of the algebra generated by
the harmonic bilocal field $V_1$. The case $d=2$}\label{apA}
\setcounter{equation}{0}

There are two main ingredients in the proof of the central result, Theorem~5.1,
of \cite{NST1} to be reviewed in the two sections of this Appendix. The first
is a (computer aided) study of the (5- and) 6-point function of the basic field
$\phi (x)$ of dimension 2, combined with the expansion (\ref{eq2.4}) (for
$\kappa = 1$) and the conservation law for the twist two fields $T_{\ell}
(x,\zeta)$ ($\ell = 2,4,\ldots$).
Here we sum up a modified version of the argument which
uses Proposition 2.1.
The (technical) difficulty of such an
analysis increases drastically with increasing the dimension of the underlying
scalar field and it has not been completed for $d=4$. The second ingredient is
quite general: it uses the discrete mode expansion of $V_1$ with respect to the
conformal Hamiltonian which can be carried out for any $d$.

\subsection{Analysis of 5- and 6-point functions for $d=2$}\label{apA1}
The general GCI and crossing symmetric 5-point function of $\phi (x)$ (for
$d_{\phi} = 2$) involves two independent terms: the sum of twelve 1-loop graphs
\begin{equation}
\label{eqA1}
w^{(1)} = \frac{c}{2} \sum_{\sigma \in {\rm Perm} (2 , \ldots , n)} (1\sigma_2)
\build{\sigma_2 \sigma_3}_{}^{\longovercorners} \ldots \build{\sigma_{n-1}
\sigma_n}_{}^{\llongovercorners} \, (1
\sigma_n) \quad \hbox{for} \ n=5
\end{equation}
where $(ij)$ is defined in (\ref{eq1.5}) and
\begin{equation}
\label{eqA2}
\build{\sigma_i \sigma_j}_{}^{\longovercorners} = (\min (\sigma_i , \sigma_j), \max
(\sigma_i ,
\sigma_j)) \, ,
\end{equation}
and a sum, $w^{(2)}$, of 10 products of 7 factors each:
\begin{equation}
\label{eqA3}
w^{(2)} = \lambda \sum_{1 \leq i < j \leq 5} \rho_{ij} \prod_{{1 \leq k \leq 5
\atop i \ne k \ne j}} \build{ik}_{}^{\overcorners} \ \build{kj}_{}^{\overcorners} \, .
\end{equation}
A rather nasty (computer aided) calculation shows that the 5-point function of three
${\phi}$'s and a $V_1$ satisfies the d'Alembert equation in the last two
arguments
\begin{equation}
\label{eqA4mew}
{\Box}_{j}\,\langle 0 \mid
{\phi}(x_1)\,{\phi}(x_2)\,{\phi}(x_3)\,V_{1}(x_4,x_5) \mid 0 \rangle = 0
\ \ \hbox{for} \ \  j = 4,\, 5 \, \
\end{equation}
iff ${\lambda}=0$. In view of Proposition 2.1 (A.4) is equivalent to
demanding the infinite set of conservation laws
\begin{equation}
\label{eqA4}
\frac{\partial^2}{\partial x_4^{\mu} \partial \zeta_{\mu}} \, \langle 0 \mid \phi
(x_1) \phi (x_2) \phi (x_3) T_{2\ell} (x_4 , \zeta) \mid 0 \rangle = 0 \, .
\end{equation}

Similarly, only the crossing symmetric sum of $\left( \frac{1}{2} \times 5 ! = 60
\right)$ 1-loop graph contributions to the truncated 6-point function is
consistent with $T_{2\ell}$ conservation. The 1-loop expression for the
6-point function allows to prove that the limit
\begin{eqnarray}
\label{eqA6}
V (x_1 , x_2) &= &\lim_{{\rho_{13} \rightarrow 0 \atop \rho_{23} \rightarrow 0}}
\{ (2\pi)^4 \rho_{13} \rho_{23} (\phi (x_1) \phi (x_2) \phi (x_3) \nonumber \\
&- &(13) \phi (x_2) - (23) \phi (x_1) - \langle 123 \rangle ) \}
\end{eqnarray}
exists, does not depend on $x_3$ and defines a harmonic in each argument bilocal
field $V_1 (x_1 , x_2)$; moreover, the truncated $n$-point functions of $\phi$
will be given by (\ref{eqA1}) for all $n$ (see \cite{NST1}, Proposition 2.3).

\medskip

{\it Remark A.1.} The above sketched analysis for the 5-point function has been
carried out in the $d=4$ case, too, with the following results. There are 37
independent GCI and crossing symmetric 5-point functions of ${\cal L}$ (compared
to 2 for $d=2$(!)). After imposing conservation of the twist 2 tensors $T_{2\ell}
(x,\zeta)$ there remain only 8, just 1 among them actually contributing to the
twist 2 part of the OPE. The latter is non-zero only if the condition
(\ref{eq4.1new}) is violated and then it yields $a_2 = 0$.

The analysis of the general 6-point function has to deal with
31990 (instead of 8 for $d=2$) independent structures
(most of them consisting of \(6!=720\) terms each).

\subsection{Analytic compact picture fields. Conformal Hamiltonian. Mode
expansions}\label{apA2}

Compactified Minkowski space
\begin{equation}
\label{eqA8}
\overline M = ({\mathbb S}^1 \times {\mathbb S}^3) / {\mathbb Z}_2
\end{equation}
(see \cite{D}) has a convenient realization in terms of (euclidean) complex
4-vectors (\cite{T2}):
\begin{equation}
\label{eqA9}
\overline M = \{ z_{\mu} = e^{2\pi i \zeta} u_{\mu} , \mu = 1,2,3,4; \zeta \in
{\mathbb R} , u \in {\mathbb S}^3 , \hbox{i.e.} \ u^2 := {\mathbf u}^2 + u_4^2 = 1 \}
\, .
\end{equation}
Real Minkowski space $M$ is mapped on an open dense subset of $\overline M$:
\begin{equation}
\label{eqA10}
M \ni (x^0 , {\mathbf x}) \rightarrow {\mathbf z} = \omega^{-1} (x) {\mathbf x} \,
, \ z_4 = \frac{1-x^2}{2\omega (x)} \, , \ \omega (x) = \frac{1+x^2}{2} - ix^0 \,
.
\end{equation}
$\overline M$ can be obtained by adding to the image of $M$ the 3-cone at
infinity:
\begin{equation}
\label{eqA11}
K_{\infty} = \{ z \in \overline M \, ; \ 1+2z_4 +z^2 (= 2(u_4 + \cos 2 \pi \zeta)
e^{2\pi i \zeta}) = 0 \} \, .
\end{equation}
Note that Eq.~(\ref{eqA10}) can be interpreted as the Cayley map from the Lie
algebra $u(2)(\simeq {\mathbb R}^4)$ to the group $U(2)$ -- cf. \cite{U}.

We note that the map (\ref{eqA10}) extends to complex arguments
\(x \mapsto x+iy\) and is regular in the forward tube (\ref{eq1.1new})
which is mapped (for any $D$) on the domain
\begin{eqnarray}
T_+ \hspace{-1pt} = \hspace{-1pt}
\left\{\raisebox{12pt}{\hspace{-3pt}}\right. z \in \mathbb{C}^D
\hspace{-2pt} : \hspace{2pt}
\left| \hspace{1pt} z^{\, 2} \right| \hspace{-1pt} < \hspace{-1pt} 1,\
2 \, \left| \hspace{1pt} z \right|^2
\left(\raisebox{12pt}{\hspace{-2pt}}\right.
= \mathop{\sum}\limits_{\mu \, = \, 1}^D \,
\left| \hspace{1pt} z_{\mu} \right|^2
\left.\raisebox{12pt}{\hspace{-2pt}}\right)
< 1 + \left| \hspace{1pt} z^{\, 2} \right|^2
\left.\raisebox{12pt}{\hspace{-3pt}}\right\}
\hspace{-2pt} .
\end{eqnarray}
where
\(z^2 = \left[ \left( 1-y^0+ix^0 \right)^2 +
\left( \mathbf{x} +i \mathbf{y} \right)^2 \right]
\left[ \left( 1+y^0-ix^0 \right)^2 +
\left( \mathbf{x} +i \mathbf{y} \right)^2 \right]^{-1}\,\).
The maximal compact subgroup $K \left( =  K \left( 4 \right) \right)$
acts by linear homogeneous transformations
on $z_{\mu}$:
$SU \left( 2 \right) \times SU \left( 2 \right)$ is
represented by real $SO \left( 4 \right)$ rotations
of the vector $\left( z_{\mu} \right)$ while
the $U \left( 1 \right)$ factor acts by phase transformations:
\(z_{\mu} \mapsto e^{i\alpha} z_{\mu}\).
Thus the point \(z=0 \in T_+\) (the image of
\(\left( x+iy \right)= \left( i,\underline{0} \right)
\in \mathfrak{T}_+\)) is left invariant by $K$.

A scalar $M$-space field $\phi_M (x)$ of dimension $d$ is related to its
$z$-picture counterpart, $\phi (z)$ (for $z=z(x)$ given in (\ref{eqA10})) by:
\begin{equation}
\label{eqA12}
\phi_M (x) = [2\pi \omega (x)]^{-d} \phi (z(x)) \, .
\end{equation}
The numerical factor $2\pi$ in (\ref{eqA12}) is chosen for convenience so that the
free massless scalar propagator assumes a simple $z$-picture form
\begin{equation}
\label{eqA13}
(12) = \frac{1}{z_{12}^2} \, .
\end{equation}

A $z$-picture scalar field $\phi (z)$ of dimension $d$ transforms in such a way
that the form $\phi (z)(dz^2)^{d/2}$ remains invariant.
(The
transformation properties of the Lagrangian ${\cal L}(z)$, for $d=4$, can
be read off, alternatively, from the invariance of the volume form
${\cal L}(z)\, dz^1 \!\wedge dz^2 \!\wedge dz^3 \!\wedge dz^4$ - cf.
(\ref{eq1.13}).)
This implies, in
particular, that correlation functions are invariant under complex euclidean
transformations. The conformal Hamiltonian $H$ (that generates translations in the
conformal time $\zeta$) acts on $\phi (z)$ and on the bilocal fields $V_{\kappa}
(z_1 , z_2)$ according to the law
$$
[H , \phi (z)] = \left( d+z \, \frac{\partial}{\partial z} \right) \phi (z) \, ,
$$
\begin{equation}
\label{eqA14}
[H , V_{\kappa} (z_1 , z_2)] = \left( 2 \kappa + z_1 \frac{\partial}{\partial z_1}
+ z_2 \frac{\partial}{\partial z_2} \right) V_{\kappa} (z_1 , z_2) \, .
\end{equation}
Here $V_{\kappa}$ are again related to $\phi$ by an expansion of type
(\ref{eq2.1}) with (12) substituted by its $z$-picture counterpart (\ref{eqA13}).

The {\it mode expansion} of $V_1 (z_1 , z_2)$ is, in particular, an expansion in
homogeneous harmonic polynomials:
$$
V_1 (z_1 , z_2) = \sum_{n,m\in {\mathbb Z}} V_{m-1 n-1} (z_1 , z_2) \, ,
$$
\begin{equation}
\label{eqA15}
V_{m-1 n-1} (\lambda_1 z_1 , \lambda_2 z_2)
= \frac{V_{m-1 n-1} (z_1 , z_2)}{\lambda_1^m \lambda_2^n} \ \hbox{for} \ \lambda_1
, \lambda_2 \in {\mathbb C}^*
\end{equation}
\begin{equation}
\label{eqA16}
\Delta_1 V_{mn} (z_1 , z_2) = 0 = \Delta_2 V_{mn} (z_1 , z_2) \ \hbox{for} \
\Delta_a = \sum_{\mu = 1}^4 \left( \frac{\partial}{\partial z_a^{\mu}} \right)^2
\, .
\end{equation}
For a hermitian scalar field $\phi$ the modes $V_{mn}$ have the following simple
conjugation property. For $m,n \geq 0$
\begin{equation}
\label{eqA17}
V_{-m-1,-n-1} (z,w) = V_{-m-1, -n-1}^{\mu_1 \ldots \mu_m \nu_1 \ldots \nu_n}
z_{\mu_1} \ldots z_{\mu_m} w_{\nu_1} \ldots w_{\nu_n}
\end{equation}
where $V_{-m-1-n-1}^{\mu_1 \ldots \mu_m \nu_1 \ldots \nu_n}$ are symmetric
traceless tensors in $\mu_1 , \ldots , \mu_m$ and in $\nu_1 , \ldots$, $\nu_n$
(separately) which are mapped under hermitian conjugation into tensors with the
same properties,
\begin{equation}
\label{eqA18}
(V_{-m-1,\, -n-1}^{\mu_1 \ldots \mu_m \nu_1 \ldots \nu_n})^* = V_{m+1 n+1}^{\mu_1
\ldots \mu_m \nu_1 \ldots \nu_n}
\end{equation}
so that
\begin{equation}
\label{eqA19}
V_{m+1 n+1} (z,w) = \frac{1}{z^2 w^2} V_{m+1 n+1}^{\mu_1 \ldots \mu_m \nu_1 \ldots
\nu_n} \frac{z_{\mu_1}}{z^2} \ldots \frac{z_{\mu_m}}{z^2} \, \frac{w_{\nu_1}}{w^2}
\ldots \frac{w_{\nu_n}}{w^2} \, .
\end{equation}

We are interested in the {\it vacuum representation} of the algebra of $V_1$ modes
for which
\begin{equation}
\label{eqA19bis}
V_{mn} \mid 0 \rangle = 0 = \langle 0 \mid V_{-m-n} \ \hbox{if either $m \geq 0$
or $n \geq 0$} \, .
\end{equation}

For $d=2$ the algebra generated by $V_{nm} (z,w)$ coincides with a {\it central
extension} $\widehat{sp} (\infty , {\mathbb R})$ {\it of the infinite symplectic
Lie algebra}. This is particularly simple to see when the (invariant under
rescaling) structure constant is a natural number:
\begin{equation}
\label{eqA20}
c := 8 \, \frac{\langle 12 \rangle \langle 13 \rangle \langle 23 \rangle
}{(\langle 123 \rangle)^2} = N \in {\mathbb N}
\end{equation}
and
\begin{equation}
\label{eqA21}
V_1 (z_1 , z_2) = \sum_{j=1}^N : \varphi_j (z_1) \varphi_j (z_2) : \quad (V_1
(z,z) = 2\phi (z)) \, ,
\end{equation}
where $\varphi_j (z)$ are free (commuting for different $j$) massless scalar fields.
The modes $\varphi_n (z)$ ($n \in {\mathbb Z}$) of each $\varphi (z)$ generate the
infinite Heisenberg algebra:
\begin{equation}
\label{eqA22}
\varphi_n (z) = e^{-2\pi i(n+1)\zeta} \varphi_n (u) \, , \ [\varphi_n (u) ,
\varphi_m (v)] = \delta_{n,-m} \, \varepsilon (n) C_{\vert n \vert - 1}^1 (uv)
\end{equation}
where $\varepsilon (n) = {\rm sign} \, n \left( = \frac{n}{\vert n \vert} \, , \
\varepsilon (0) = 0 \right)$ $C_n^1$ are the Gegenbauer polynomials generated by
the 2-point function of $\varphi$:
$$
C_{n-1}^1 (\cos 2 \pi \alpha) = \frac{\sin 2 \pi n \alpha}{\sin 2 \pi \alpha} \, ,
\ \frac{1}{z_{12}^2} = \frac{1}{z_1^2} (1-2xz+z^2) = \frac{1}{z_1^2}
\sum_{n=0}^{\infty} z^n C_n^1 (x) \, ,
$$
\begin{equation}
\label{eqA23}
z = \sqrt{\frac{z_2^2}{z_1^2}} \, , \ x = uv \, .
\end{equation}
It is well known that the quadratic combination of the generators of a Heisenberg
algebra give rise to a symplectic Lie algebra. The central extension comes, as
usual, from the normal products.

The algebra $\widehat{sp} (\infty , {\mathbb R})$ of $V_{nm} (u,v)$ has a simple
{\it diagonal subalgebra}, generated by $v_{nm} := V_{nm} (u,u)$, $u \in {\mathbb
S}^3$,
\begin{eqnarray}
\label{eqA24}
[v_{n_1 m_1} , v_{n_2 m_2}] &= &c \, n_1 m_1 (\delta_{n_1 , -n_2} \delta_{m_1 , -m_2}
+ \delta_{n_1 , -m_2} \delta_{m_1 , -n_2}) \nonumber \\
&+ &n_1 (\delta_{n_1 , -n_2} v_{m_1 m_2} + \delta_{n_1 , -m_2} v_{m_1 n_2})
\nonumber \\
&+ &m_1 (\delta_{m_1 , -n_2} v_{n_1 m_2} + \delta_{m_1 , -m_2} v_{n_1 n_2}) \, .
\end{eqnarray}
The proof of Theorem~5.1 of \cite{NST1} is now based on the construction of a
sequence $\langle \Delta_n \mid$, $n = 1,2, \ldots$ of vectors (Lemma~5.2)
$$
\langle \Delta_n \mid = \frac{1}{n!} \langle 0 \mid \left\vert \matrix{
v_{11} &v_{12} &\ldots &v_{1n} \cr
v_{21} &v_{22} &\ldots &v_{2n} \cr
\ldots &\ldots &\ldots &\ldots \cr
v_{n1} &v_{n2} &\ldots &v_{nn} \cr
} \right\vert
$$
\begin{equation}
\label{eqA25}
\hbox{such that} \ \langle \Delta_n \mid \Delta_n \rangle = (n+1)! c(c-1) \ldots
(c-n+1) \, ,
\end{equation}
implying $c \in {\mathbb N}$ for unitary vacuum representations of $\widehat{sp}
(\infty , {\mathbb R})$. Then one deduces that Wightman positivity implies that
$V_1$ has the form (\ref{eqA21}) for some $N$.

The difficulty in extending this analysis to the case $d=4$ again lies in the
necessity of having the 6-point function of ${\cal L}$ in order to be able to
compute the structure constants of the infinite Lie algebra generated by the modes
of $V_1 (z_1 , z_2)$.

\end{document}